\documentclass{aa}

\usepackage{graphicx}
\usepackage{rotating}
\usepackage{txfonts}
\usepackage{color}
\usepackage{natbib}
\bibpunct{(}{)}{;}{a}{}{,} 

\def\Msun{$M_{\odot}$}

\def\Teff{\ensuremath{T_{\mathrm{eff}}}}
\def\cd{d$^{\rm -1}$}
\def\logg{\ensuremath{\log g}}

\def\vsini{\ensuremath{{\upsilon}\sin i}}
\def\kms{$\mathrm{km\,s}^{-1}$}

\def\espa{ESPaDOnS}

\begin{document}

\title{A comprehensive study of young B stars in NGC 2264: \\I. Space photometry and asteroseismology \thanks{Based on data from the {\it MOST} satellite, a Canadian Space Agency mission, jointly operated by Microsatellite Systems Canada Inc. (MSCI), formerly part of Dynacon, Inc., the University of Toronto Institute for Aerospace Studies and the University of British Columbia with the assistance of the University of Vienna. Based on observations obtained at the Canada-France-Hawaii Telescope (CFHT), which is operated by the National Research Council of Canada, the Institut Naional des Sciences de l'Univers of the Centre National de la Recherche Scientifique of France, and the University of Hawaii.}}

\author{K. Zwintz\inst{1}\thanks{Elise Richter fellow of the Austrian Science Funds (FWF)} \and
E. Moravveji\inst{2}\thanks{Marie Curie Postdoctoral Fellow} \and
P.~I.~P\'{a}pics\inst{2}\thanks{Postdoctoral Fellow of The Research Foundation -- Flanders (FWO), Belgium} \and 
A. Tkachenko\inst{2} \and
N. Przybilla\inst{1} \and
M.-F. Nieva\inst{1} \and
R. Kuschnig\inst{3} \and
V. Antoci\inst{4} \and
D. Lorenz\inst{5} \and
N. Theme{\ss}l\inst{6,4} \and 
L. Fossati\inst{7} \and
T. G. Barnes\inst{8}
}

\offprints{K. Zwintz \\ \email{konstanze.zwintz@uibk.ac.at}}

\institute{
   Institut f\"ur Astro- und Teilchenphysik, Universit\"at Innsbruck,  Technikerstrasse 25, A-6020 Innsbruck, Austria \\
    \email konstanze.zwintz@uibk.ac.at  \and
Instituut voor Sterrenkunde, KU Leuven, Celestijnenlaan 200D, 3001, Leuven, Belgium \and
 Graz University of Technology, Institute for Communication Networks and Satellite Communication, Inffeldgasse 12, A-8010 Graz \and
Stellar Astrophysics Centre, Department of Physics and Astronomy, Aarhus University, Ny Munkegade 120, 8000, Aarhus C, Denmark \and
University of Vienna, Institute for Astrophysics, T\"urkenschanzstrasse 17, A-1180 Vienna, Austria \and
Max Planck Institute for Solar System Research, Justus-von-Liebig-Weg 3, D-37077 G\"ottingen, Germany \and 
Space Research Institute, Austrian Academy of Sciences, Schmiedlstrasse 6, A-8042 Graz, Austria \and
The University of Texas at Austin, McDonald Observatory, 2515 Speedway, Stop C1402, Austin, Texas 78712, USA
    }

\date{Received / Accepted }

\abstract
{Space photometric time series of the most massive members of the young open cluster NGC 2264 allow us to study their different sources of variability down to the millimagnitude level and permits a search for Slowly Pulsating B (SPB) type pulsation among objects that are only a few million years old.}
{Our goal is to conduct a homogeneous study of young B type stars in the cluster NGC 2264 using photometric time series from space in combination with high-resolution spectroscopy and spectropolarimetry obtained from the ground. The latter will be presented in a separate follow-up article.}
{We performed frequency analyses for eleven B stars in the field of the young cluster NGC 2264 using photometric time series from the MOST, CoRoT and Spitzer space telescopes and the routines {\sc Period04} and {\sc SigSpec}. We employ the MESA stellar evolution code in combination with the oscillation code GYRE to identify the pulsation modes for two SPB stars which exhibit short period spacing series.}
{From our analysis we identify four objects that show SPB pulsations, five stars that show rotational modulation of their light curves caused by spots, one star that is identified to be a binary, and one object in the field of the cluster that is found to be a non-member Be star. In two SPB stars we detect a number of regularly spaced pulsation modes that are compatible with being members of a g mode period series. }
{Despite NGC 2264's young age, our analysis illustrates that its B type members have already arrived on the zero-age main sequence (ZAMS). Our asteroseismic analysis yields masses between 4 and 6 \Msun\, and ages between 1 and 6 million years, which agree well to the overall cluster age. }


\titlerunning{Young B stars in NGC 2264}
\authorrunning{K. Zwintz et al.}
\maketitle

\section{Introduction}

Stellar rotation, pulsation, binarity and magnetic fields play important roles in massive stars influencing their structure, evolution and lifetimes across all evolutionary phases, from star formation until the latest stages \citep[e.g.,][]{lan12,mae12,aer10}. Our theoretical concepts for the evolutionary properties of O and B stars have strong implications on the predictions for supernovae progenitors, galaxies and the chemical enrichment of the whole universe.
Stellar evolution codes of single and binary stars include parametrized descriptions of phenomena crucial for massive stars, such as internal differential rotation, transport of angular momentum, convective core overshooting, chemical composition, and magnetic fields. However, our theoretical concepts mostly lack essential calibration by high-precision observations. First successful seismic studies revealed, for example, the internal rotation profile \citep{tri15} or put tight constraints on core overshooting and diffusive mixing of a massive pulsating star based on observations conducted with the {\it Kepler} space telescope \citep{bor10}, hence illustrating the importance and power of testing our theoretical predictions with precise observational data.
 
Slowly pulsating B (SPB) stars \citep{wae91} have spectral types between B3 and B9 corresponding to $\sim$ 11\,000 to 22\,000\,K in effective temperature (\Teff) and masses ranging from 2.5 to 8\,\Msun\ \citep[see e.g.,][Chapter 2]{aer10}. They show non-radial, multi-periodic, high-order gravity (g) mode oscillations  with periods between 0.5 to 3 days which are driven by the $\kappa$-mechanism connected to the iron-group elements opacity bump \citep{dzi93}. The SPB stars' g-modes of same degree $\ell$ and consecutive radial order $n$ are expected to be nearly equally spaced in period. The deviations from this equal spacing carry information about the physical processes occurring near the core \citep{mig08}. 

This theoretical prediction was first investigated with observational data by \citet{pap15}, who detected a long series of g-modes with nearly equidistant spacings in period space for an SPB star which revealed clear signatures of chemical mixing and rotation.
Moreover, asteroseismic modeling allowed ages for several SPB stars to be deduced from their central hydrogen mass fractions, $X_c$ \citep[e.g.,][]{mor15,mor16}, always under the assumption that the stars are in their main sequence evolutionary phases, i.e., burning hydrogen in their cores. 

\citet{gru12} suggested for the first time the presence of SPB stars among pre-main sequence (pre-MS) members of the young open cluster NGC 2244 based on an asteroseismic analysis. 
Stars with masses higher than $\sim$6\Msun\ do not have an optically observable pre-MS phase as the birthline intersects the ZAMS \citep{pal90,pal93}. B type stars with masses lower than $\sim$6\Msun\ have a relatively short optically visible pre-MS evolutionary phase lasting at most a few million years. Generally, pre-MS stars mainly gain their energy from gravitational contraction until their cores become hot enough for hydrogen burning to ignite. As pre-MS SPB stars are already quite close to their arrival on the ZAMS, they are in a transition phase between the end of gravitational contraction as main energy source and the onset of hydrogen core burning. During this phase, stars undergo significant structural changes that shall be probed through asteroseismology. Only with the first observations of potential pre-MS SPB candidates from space using the MOST satellite \citep{wal03}, the search for and the investigation of this type of young pulsating objects could commence.
For a full interpretation of the stars' photometric variability, for any asteroseismic interpretation and for tests of stellar evolution models, an accurate characterization of their fundamental parameters such as effective temperature, surface gravity, chemical abundances and magnetic fields is crucial. 

The present study is the first of two articles prepared simultaneously that use the time series photometry for eleven B stars in the field of NGC 2264 in combination with high-resolution spectroscopy and spectropolarimetry for a comprehensive study of the young B stars in NGC 2264. In this paper (Paper I hereafter) we discuss the space photometry obtained with the MOST space telescope \citep{wal03} in 2006 and 2011/12, with the CoRoT satellite \citep[COnvection, ROtation and Planetary Transits;][]{bag06} in 2008 and with the NASA Spitzer \citep{wer04} satellite obtained during the CSI2264 (Coordinated Synoptic Investigation of NGC 2264) project \citep{cod13} simultaneously to the 2011/12 MOST (and CoRoT) observations (Sect. \ref{photometry}). We characterize the different types of variability (Sect. \ref{photometry}), conduct an asteroseismic investigation for two selected young SPB stars showing indications for period spacings, and discuss the stars' evolutionary stage (Sect. \ref{sec-models}).

The subsequent paper (Przybilla et al. 2017, in preparation; Paper II hereafter) will focus on the analysis of the high-resolution spectroscopy and spectropolarimetry obtained with the Mc Donald Observatory 2.7\,m telescope and the Tull spectrograph  and with \espa\,\, at the Canada-France-Hawaii Telescope (CFHT) and describe the stars' fundamental parameters, atmospheric chemical abundances, and magnetic properties.

\section{B stars in the field of NGC 2264} \label{star_description}

NGC 2264 ($\alpha_{\rm 2000}$ = 06${\rm ^h}$\,41${\rm ^m}$, $\delta_{\rm 2000}$ = +09$^{\circ}$\,35') is a very young \citep[age $\sim$3 -- 8 Myr; e.g.,][]{sun04,sag86} open cluster that is rich in young stellar objects and has been studied frequently in the past using various instruments in different wavelength ranges from space and from the ground. It has a diameter of $\sim$39 arcmin and is located in the Monoceros OB1 association about 30pc above the Galactic plane. The cluster distance is reported to be 759 $\pm$ 83 pc by \citet{sun97} corresponding to a distance modulus of 9.40 $\pm$ 0.25 mag. The proper motion values of NGC 2264 are $-2.70 \pm 0.25$ mas/yr in right ascension and $-3.50 \pm 0.26$ mas/yr in declination \citep{kha01}. The cluster reddening is rather low with $E(B-V) = 0.071 \pm 0.033$ mag and the differential reddening is negligible \citep{sun97}. 

NGC 2264 is located in a region accessible to the space telescopes MOST, CoRoT, Spitzer and Chandra and was the target for the Coordinated Synoptic Investigation of NGC 2264 (CSI2264) project \citep{cod13}. The CSI2264 project lasted from the beginning of December 2011 until early January 2012 and involved simultaneous observations of NGC 2264 for more than one month using all four above mentioned space telescopes together with a coordinated ground-based observing campaign for complementary data. 
One of the main scientific topics addressed with these combined observations from space and from the ground was the discovery and analysis of pulsationally variable stars in the pre- and early main sequence stages. 
\citet{zwi09} presented the first investigation of 68 stars in the field of NGC 2264 observed by the MOST satellite in 2006. They identified four pre-main sequence (pre-MS) $\delta$ Scuti type stars, one $\gamma$ Doradus candidate, ten SPB candidates, and six stars showing rotational modulation caused by spots on their surfaces.  
A previous investigation of the two early B-type stars HD 47887 and HD 47777 in NGC 2264 revealed that both stars possess large-scale magnetic fields with average longitudinal field strengths of about 400\,G  \cite{fos14}.  For completeness of the sample of B stars in NGC 2264 observed from space, we also include these two objects in the present study. In the following investigation we will refer to the results published in that paper and put both stars into context to the nine other stars. 

\begin{figure}[htb]
\centering
\includegraphics[width=0.48\textwidth,clip]{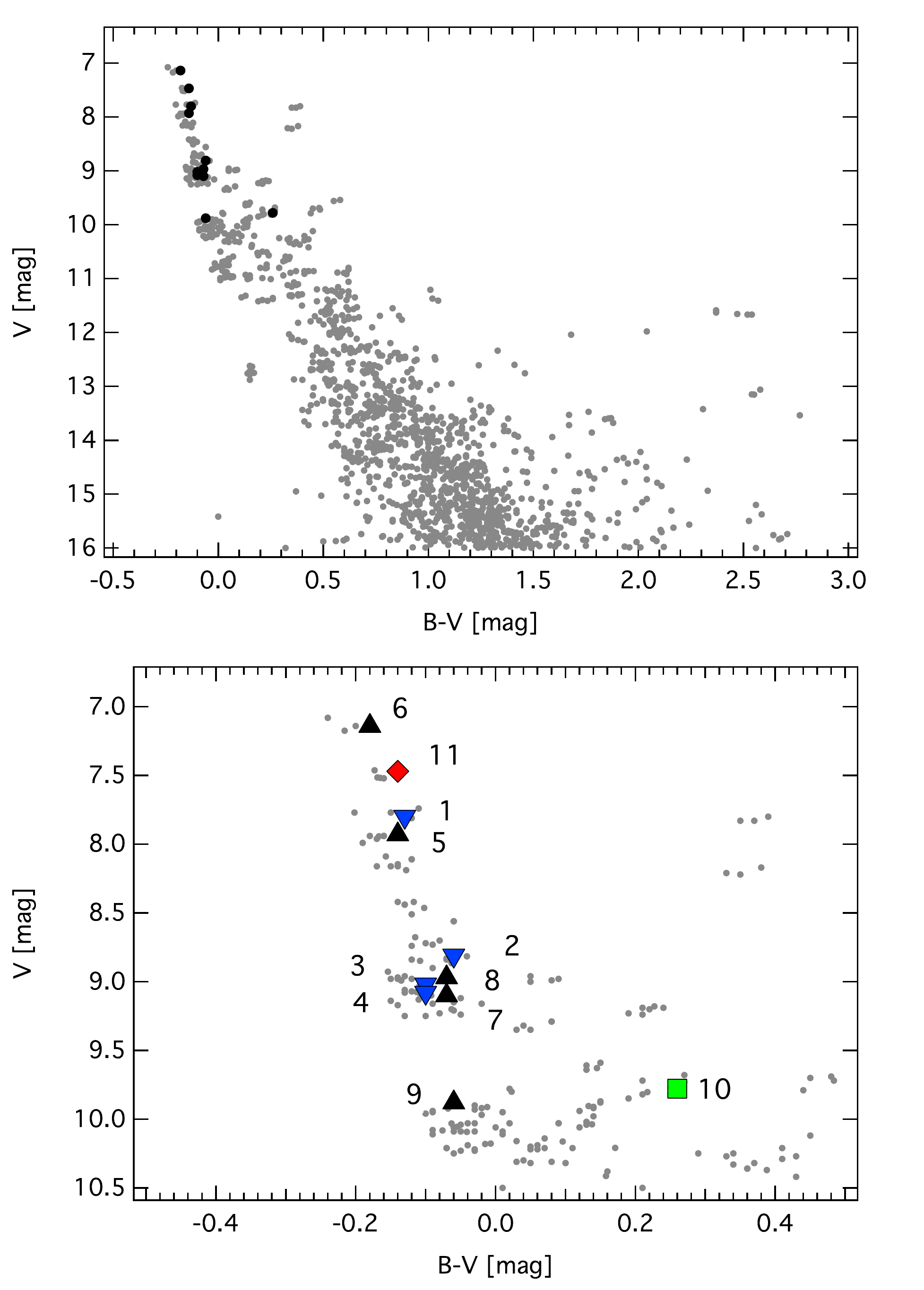}
\caption{Colour-magnitude diagram (CMD) of the open cluster NGC 2264 in the Johnson $B$ and $V$ filters \citep[grey points, source: WEBDA database of open clusters;][]{mer03}: top panel -- complete cluster CMD where the eleven stars are marked as black symbols; bottom panel -- zoom into the most luminous part of the CMD where the pulsators are marked with blue downwards triangles, stars showing rotational modulation are identified as black upward triangles, the Be star is shown as green square and the binary is given as a red diamond. Numbers are according to the identification in Table \ref{sample}.}
\label{cmd}
\end{figure}

\begin{figure}[htb]
\centering
\includegraphics[width=0.4\textwidth,clip]{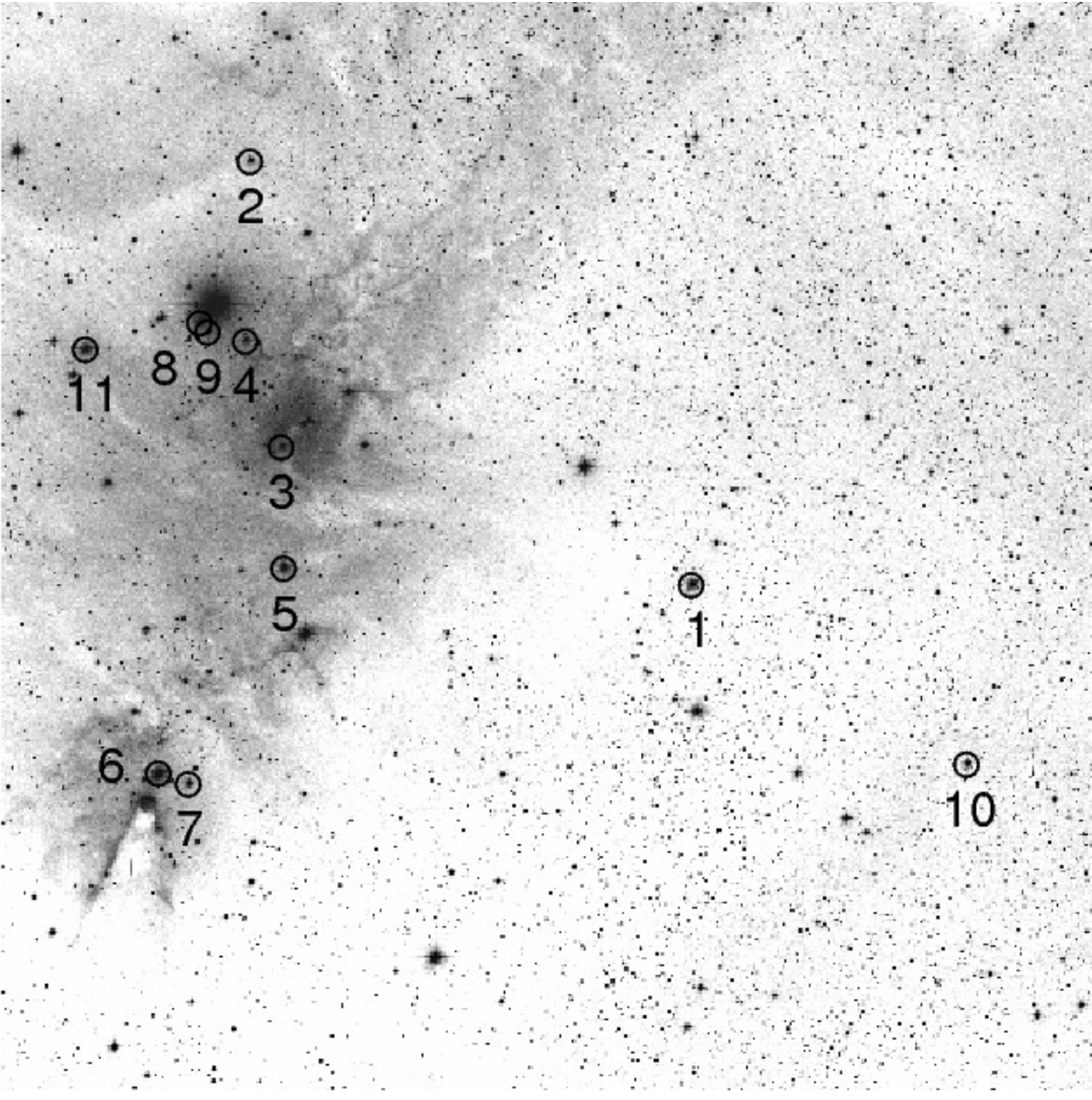}
\caption{ DSS image of a 1 x 1 degree$^2$ region centered on RA$_{\rm 2000}$ = 06:39:43.5 and DE$_{\rm 2000}$ = 09:40:50.3 showing the positions of the eleven B stars in the sky. Numbers are given as in Table \ref{sample}.}
\label{sky}
\end{figure}

The stars HD 47469, HD 48012, HD 261810, HD 261878, HD 47777, HD 47887, HD 261903, HD 261938, NGC 2264 137, HD 261054 and HD 47961 (numbered from 1 to 11 accordingly; see also Table \ref{sample}) lie in the magnitude range between 7.1 and 9.9 mag in the $V$ band, and -- given their corresponding smallest $B-V$ values -- they belong to the hottest stars in NGC 2264. 
Note that the spectral type of "K1?'' given by the SIMBAD data base for HD 261903 was derived from multi-color photometry and is likely a mis-identification: As HD 261903 is located in a relatively obscured region in NGC 2264 (which can also be identified in Figure \ref{sky}), its colors would indeed point to a cool object, but its spectrum clearly indicates a spectral type B.
The top panel of Fig. \ref{cmd} shows the colour-magnitude diagram (CMD) of NGC 2264 in Johnson $V$ and $B$ photometry down to $V=16$ mag, where the eleven B type stars are marked with black symbols. A zoom into the hottest part of the CMD (Fig. \ref{cmd} bottom panel) makes it easier to identify the stars that are subject to this investigation. 

All stars, except for NGC 2264 137, have been reported to be candidates for SPB type pulsations showing frequencies between 0.05\cd\ and 6.5\cd\ by \citet{zwi09}, but the authors did not describe the frequency content in detail. NGC 2264 137 (GSC 00750-01804) did not show clear variability in the MOST 2006 observations, hence was listed as a constant object \citep{zwi09}. 
The star HD 48012 is listed in the CHANDRA Variable Guide Star Catalog as a pulsating variable star and UV source \citep{nic10}.
The stars HD 261810, HD 261903, HD 261938 and HD 47961 are reported to be X-ray sources from ROSAT measurements \citep{fla00} and HD 261810 and NGC 2264 137 are included in the X-ray observations by \citet{dah07}.
The stars HD 47777 and HD 47887 have been reported to show variability caused by rotational modulation of a spotted surface and host large scale magnetic fields \citep{fos14}.

For HD 261810, HD 261878, HD 261903, HD 261938, NGC 2264 137 and HD 47961 membership probabilities (see Table \ref{sample}) ranging between 25\% (for HD 261903) and 97\% (for HD 261938) were taken from \citet{vas65} and \citet{fla00}. All stars were also included in the CSI2264 project where for HD 47469 and HD 261054 no membership information is given and for all other stars a very likely to possible cluster membership is reported (see Table \ref{sample}). From radial velocity measurements, \citet{fos14} concluded that HD 47777 and HD 47887 are cluster members.

Four additional criteria can be used to assess the cluster membership of stars: (i) their position on the sky with respect to the center of the cluster, (ii) their proper motion values in comparison to the cluster proper motion, (iii) the location of the stars in the cluster colour-magnitude diagram (CMD), and (iv) the stars' radial velocities.
The location of the eleven stars on the sky is displayed in Fig. \ref{sky}, which shows a one square degree field centered at the coordinates RA$_{\rm 2000}$ = 06:39:43.5 and DE$_{\rm 2000}$ = 09:40:50.3. It can clearly be seen that the two stars labeled \#1 (HD 47469) and \#10 (HD 261054) are located further away from the cluster center compared to the other nine objects. 
However, HD 47469's CMD position matches quite well those of the other stars with high membership probability (e.g., NGC 2264 137 with 96\%) in its vicinity. Additionally, its proper motion agrees to the cluster values within the given errors (Table \ref{sample}). Hence, we find more arguments to support HD 47469's membership to NGC 2264 than against it.
On the other hand, HD 261054, is positioned with an offset in $(B-V)$ in the cluster CMD (Fig. \ref{cmd}) and has a proper motion that is significantly outside the errors of the cluster values. We therefore conclude, that HD 261054 is more likely to be a background field object.
The radial velocities of all eleven stars have been determined homogeneously from our spectroscopic data and will be discussed in detail in Paper II (Przybilla et al. 2017).

Four stars of our sample have also entries in the recently published Gaia Data Release 1 \citep[DR1,][]{gaia16}: HD 48012 with a parallax of 1.48 $\pm$ 0.36 mas, HD 261810 1.53 $\pm$ 0.41 mas, HD 261878 with 1.48 $\pm$ 0.41 mas and HD 261054 with 0.75 $\pm$ 0.23 mas corresponding to distances of 675$^{+133}_{-220}$pc, 652$^{+136}_{234}$pc, 675$^{+147}_{-261}$pc and 1342$^{+321}_{-616}$pc. These results confirm our conclusion that HD 261054 is not a member of NGC 2264 while the values for the other three objects basically agree to the mean cluster distance.

Hence, we conclude that in our sample ten stars are likely to be cluster members and only HD 261054's membership to NGC 2264 is unlikely.

\begin{table*}[htb]
\caption{Properties of the eleven B stars in the field of NGC 2264 and comparison to the cluster position and proper motions \citep{kha01}. }
\label{sample}
\begin{center}
\begin{scriptsize}
\begin{tabular}{rlrccccrrccc}
\hline
\hline
\multicolumn{1}{l}{\#} &\multicolumn{1}{l}{Name} &\multicolumn{1}{l}{Mon-ID}& \multicolumn{1}{c}{RA [2000.0]} & \multicolumn{1}{c}{DE [2000.0]} & \multicolumn{1}{c}{$V$} & \multicolumn{1}{c}{sp} & \multicolumn{1}{c}{pmRA}  & \multicolumn{1}{c}{pmDE} & \multicolumn{3}{c}{membership}   \\
\multicolumn{1}{l}{ } & \multicolumn{1}{l}{ } &\multicolumn{1}{l}{ } & \multicolumn{1}{c}{[hh:mm:ss]} & \multicolumn{1}{c}{[dd:mm:ss]} & \multicolumn{1}{c}{ [mag]} & \multicolumn{1}{l}{ [SIMBAD]} & \multicolumn{1}{c}{[mas/yr]}  & \multicolumn{1}{c}{[mas/yr]}  & \multicolumn{1}{c}{prob. }  & \multicolumn{1}{c}{ $Ref$ }& \multicolumn{1}{l}{ CSI2264}  \\
\hline
1 &	HD 47469	 & Mon-024786	&	 06:39:11.172 &  	+09:38:45.47	 		& 	7.80 & O/B2		&	$-3.4 \pm 1.9$	&	$-1.2 \pm 1.9$	& - 	& -  & 0	\\
2 &	HD 48012	 & Mon-000030	&	06:40:51.146 &		+10:01:43.93			&	8.81	 & B6/9	&	$-0.6 \pm 1.1$	&	$-4.2 \pm 1.2$	& - 	& - & 1	\\
3 &	HD 261810 & Mon-000523	&	06:40:43.227 &		+09:46:01.71			&	9.02	 & B1	&	$0.0 \pm 1.9$	&	$-4.9 \pm 1.9$	& 74\% 	& 1, 2 & 1	\\
4 &	HD 261878 & Mon-005579	&	06:40:51.558 &		+09:51:49.39			& 	9.08	& B6		&	$-0.1 \pm 1.6$	&	$-2.5 \pm 1.5$	& 94\%	& 1  & 2 \\
5 & HD 47777 & Mon-000248		& 06:40:42.293 			& +09:39:21.31			&	7.93 & B3		&	$-0.4 \pm 1.9$	&	$-5.5 \pm 1.9$	&	98\%		&	1	& 1 \\
6 & HD 47887 & Mon-000303		&	06:41:09.602 &	 +09:27:57.54	&		7.14	&		B2		&	$1.5 \pm 1.5$	&	$-1.9 \pm 1.5$	&	7\%		&	1	&	1 \\
7 &	HD 261903 & Mon-000193	& 	06:41:02.886 &		 +09:27:23.49			&	9.10	 & K1?$\ast$	& 	$-0.1 \pm 2.0$	&	$-3.9 \pm 1.9$	& 25\%	& 1  & 1\\
8 &	HD 261938 & Mon-000270	&	06:41:01.874 &		+09:52:47.96 			& 	8.97	& B3		& 	$-7.5 \pm 2.0$	&	$-12.9 \pm 1.8$ 	& 97\%	& 1,2 & 1  \\
9 &	NGC 2264 137 & Mon-000902	& 06:41:00.03 &		 +09:52:17.9			&	9.88	& A1 	&	$-4.0 \pm 5.0$	&	$-3.4 \pm 1.0$	& 96\% 	& 1, 2 & 1	 \\
10 & 	HD 261054 & Mon-052766	& 	06:38:09.175 &		+09:29:05.24			&	9.78	 & O/B2	&	$+2.2 \pm 2.6$	&	$+0.2 \pm 2.5$	&  -	& - & 0	\\
11 &	HD 47961	& Mon-000068	&	06:41:27.304 &		 +09:51:14.41			&	7.47	 & B2	&	$-1.3 \pm 1.4$	&	$-5.2 \pm 1.3$	& 96\% 	& 1, 2  & 1 \\
\hline
   &   NGC 2264 &	&	06:40:58.0	&	+09:53:42.0			&		&		&	$-2.70 \pm 0.25$	&	$-3.50 \pm 0.26$ & & & \\
\hline
\end{tabular}
\end{scriptsize}
\end{center}
\tablefoot{Column headers are a running number (\#), identifier (Name), CSI2264 project identifier (Mon-ID), right ascension and declination to the epoch of 2000.0 (RA, DE), $V$ magnitude, spectral type as listed by SIMBAD (sp), proper motions in RA and DE (pmRA, pmDE) taken from the TYCHO2-Catalogue \citep{hog00} except for NGC 2264 137 which is taken from the ASCC-2.5 V3 Catalogue \citep{kha01}, membership probability (prob.) as given in the literature listed in the column $Ref$ where Reference \# 1: \citet{vas65}; reference \# 2: \citet{fla00} and membership as provided from the CSI2264 project through the NASA / IPAC Infrared Science Archive\footnote{http://irsa.ipac.caltech.edu/} (CSI2264), where 0 -- no membership information, 1 -- very likely cluster member, 2 -- possible cluster member. Cluster proper motion values are taken from \citet{kha01}. $\ast$: The spectral type of K1? for HD 261903 is likely a misidentification based on multi-color photometry of a star located in an obscured region of the cluster. }
\end{table*}

\section{Photometric analysis}
\label{photometry}

The photometric time series presented in this study have been obtained using the three space telescopes MOST, CoRoT and Spitzer. 
Table \ref{obs} summarizes the observational material.

\subsection{MOST observations and data reduction}

The Canadian micro-satellite MOST carries a 15-cm Rumak Maksutov telescope which feeds a CCD photometer through a single custom broadband filter in the 3500--7500\AA\,\, wavelength range. Three types of photometric data can be supplied simultaneously for multiple targets in a given field: Fabry Imaging, Direct Imaging and Guide Star Photometry data. Although the Canadian Space Agency ceased funding for MOST in 2015, the space telescope is still operational and can be used for dedicated observing runs that can be purchased by the scientific community.

\begin{table}[htb]
\caption{Overview of the photometric observations from space. }
\label{obs}
\begin{center}
\begin{scriptsize}
\begin{tabular}{rlcccc}
\hline
\hline
\multicolumn{1}{l}{\#} &\multicolumn{1}{l}{Name} & \multicolumn{1}{c}{MOST 2006} & \multicolumn{1}{c}{MOST 2011/12} & \multicolumn{1}{c}{other} & \multicolumn{1}{c}{var}\\
\multicolumn{1}{l}{ } & \multicolumn{1}{l}{ } & \multicolumn{1}{c}{GS\# / field} & \multicolumn{1}{c}{GS\# / field} & \multicolumn{1}{c}{ } & \multicolumn{1}{c}{ }\\
\hline
1 &	HD 47469		&	04 / B	& 	04 / 1		&	 & SPB 	\\
2 &	HD 48012		&	14 / A	& 	-			&	 & SPB	\\
3 &	HD 261810	&	21 / B	&	13 / 1		&		 & SPB	\\
4 &	HD 261878	&	28 / B	&	23 / 1		&	 & SPB \\
\hline
5 &	HD 47777		&	18 / B	&	12 / 1		&		&	rot. mod. \\
6 &	HD 47887		&	19 / A	&	-			&		& rot. mod. \\
7 &	HD 261903 	& 	13 / A	&	18 / 2	&	Spitzer		 & rot. mod. \\
8 &	HD 261938	&	20 / A	& 	28 / 1 		&	 & rot. mod. \\
9 &	NGC 2264 137 &	18 / A	&	27 / 1	&  CoRoT		& rot. mod. \\
\hline
10 & 	HD 261054	& 	00 / B	&	-			&	 & Be	\\
\hline
11 &	HD 47961		&	31 / A	& 	38 / 1 	&	 & SB \\
   &					&			&	19 /2   &    &  \\
\hline
\end{tabular}
\end{scriptsize}
\end{center}
\tablefoot{Column headers are a running number as in Table \ref{obs} (\#), identifier (Name), MOST guide star numbers and fields in 2006 (MOST 2006) and 2011/12 (MOST 2011/12), availability of additional CoRoT or Spitzer data (other), and identified type of variability (var): slowly pulsating B type (SPB), rotational modulation (rot.mod), Be star (Be) and spectroscopic binary (SB). }
\end{table}

MOST observed the young open cluster NGC~2264 for the first time from December 7, 2006, to January 3, 2007, in a dedicated observing run on the cluster itself \citep{zwi09}. 
The second MOST observing run on NGC~2264 lasted from December 5, 2011, to January 14, 2012, and was conducted as part of the CSI 2264 project.

Due to the magnitude range (7 $< V <$ 12 mag) and the large number of targets, in both runs NGC~2264 was observed in the open field of the MOST Science CCD in Guide Star Photometry Mode. Because not all of the targets could be reached using a single pointing of the satellite, two fields of observations were chosen and observed in alternating halves of each 101-min orbit. 
Using this setting, MOST time series photometry was obtained for a total of 68 stars in the region of NGC~2264 in the 2006 run \citep{zwi09} and 67 in the 2011/12 run.

\begin{figure*}[htb]
\centering
\includegraphics[width=0.8\textwidth,clip]{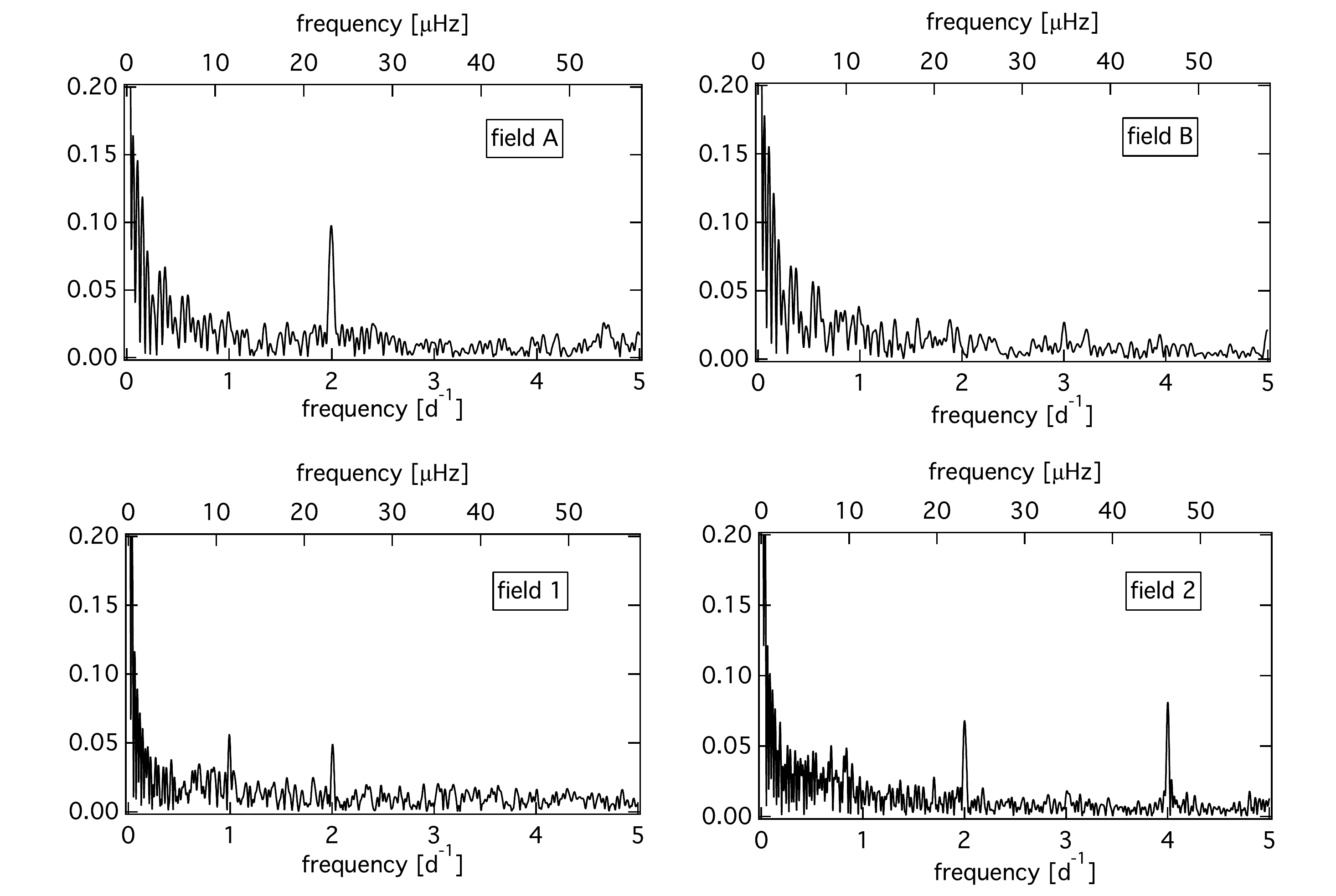}
\caption{Spectral windows obtained from MOST data 2006 fields A and B (top panels) and 2011/12 fields 1 and 2 (bottom panels). Note the necessary zoom in the Y-axis to make some of the structure visible.}
\label{hd47469spws}
\end{figure*}

Data reduction of the MOST Guide Star photometry was conducted using the method developed by \cite{har08} and is described in detail in \citet{zwi09}. 
The corresponding MOST 2006 light curves of both fields - field A and field B - used for the analysis each have a time base of $\sim$22.7\,d. The on-board exposure time was 1.5\,s, 16 consecutive images were ``stacked" on board resulting in an integration time of 24\,s \citep[see also][]{zwi09}.
The total length of the NGC 2264 data set from the 2011/12 run is 38.77\,d for field 1 and 39.98\,d for field 2. In 2011/12  on-board exposures were 3.01\,s long (to satisfy the cadence of guide star ACS operations), and the integration time was 51.17\,s as 17 consecutive images were ``stacked" on board.
An overview of the properties of the MOST data is given in Table \ref{most}. 
The corresponding spectral windows (top panels: 2006 data, bottom panels: 2011/12 data) are given in Fig. \ref{hd47469spws}. 

\begin{table}[htb]
\caption{Characteristics of the MOST, CoRoT and Spitzer data sets. }
\label{most}
\begin{center}
\begin{scriptsize}
\begin{tabular}{lcrrcr}
\hline
\hline
\multicolumn{1}{l}{data set} & \multicolumn{1}{c}{obs. info}& \multicolumn{1}{c}{N} & \multicolumn{1}{c}{t$_{base}$} & \multicolumn{1}{c}{$1/T$} & \multicolumn{1}{c}{f$_{\rm Nyquist}$}  \\
\multicolumn{1}{c}{ } & \multicolumn{1}{c}{ } & \multicolumn{1}{c}{\#}  & \multicolumn{1}{c}{[d]} & \multicolumn{1}{c}{[d$^{-1}$]} & \multicolumn{1}{c}{[d$^{-1}$]} \\
\hline
MOST06 & field A & 10531 & 22.72 & 0.044 &  1362.23 \\
  & field B & 17698 & 22.72 & 0.044 &  1416.73 \\
MOST11/12 & field 1 & 16688 & 38.77 & 0.026 & 721.34  \\
  & field 2 & 8371 & 39.98 & 0.025 & 705.40  \\
CoRoT08 & SRa01 & 46780 & 19.28 & 0.052 & 1338.15 \\
Spitzer11/12 & staring mode & 18258 & 6.01 & 0.166 & 2769.22 \\
\hline
\end{tabular}
\end{scriptsize}
\end{center}
\tablefoot{Data set, information on the observations (obs. info), number of data points used for the analysis ($N$), time base (t$_{base}$), Rayleigh frequency resolution ($1/T$), and Nyquist frequency (f$_{\rm Nyquist}$).}
\end{table}

\subsection{CoRoT observations and data reduction}

On December 27, 2006, the European satellite CoRoT was launched from Baikonur, Kazakhstan, into a circular polar orbit of about 896 km altitude. 
It carried a 27-cm telescope and observed stars inside two cones with a radius of 10$^{\circ}$ centered around RA\,=\,06:50 and RA\,=\,18:50. The telescope's field-of-view is almost circular with a diameter of 3.8$^{\circ}$. The filter bandwidth ranges from 370 to 1000\,nm.
The CoRoT space telescope originally had two CCDs devoted to asteroseismology for stars with $5.7 < m_V < 9.5$ mag and two CCDs dedicated to the search for exoplanets where $\sim$6000 stars in the magnitude range from 10 to 16 mag in $R$ per CCD were monitored simultaneously.
Since a hardware failure in November 2012, CoRoT is not operational any more. 

The young cluster NGC 2264 was targeted by CoRoT for the first time during the Short Run SRa01 in 2008 for 23.4 days consecutively and for the second time during the Short Run SRa05 in 2011/12 as part of the previously mentioned CSI 2264 multi-satellite campaign for a total time base of $\sim$39 days. In both observing runs, the complete cluster was placed in one exofield CCD and data were taken for all stars in the accessible magnitude range.

Therefore, the brightest stars in NGC 2264 (i.e., with $V < 10$ mag) have not been targets for the CoRoT exofield observations. But for selected stars data could be obtained using the method of the so-called CoRoT imagettes \citep[for a detailed description see, e.g.,][]{zwi11}. NGC 2264 137 is the only B-type star in the field of the cluster that has been included as an imagette in the CoRoT SRa01 observations of NGC 2264. Due to the special observations of a star that is nominally too bright and saturates on the CCD, the light curve of NGC 2264 137 is only $\sim$ 19 days long (see Table \ref{most}).

\subsection{Spitzer observations}
During the CSI2264 campaign, the {\it Spitzer} satellite observed the young cluster NGC 2264 for roughly 30 days in the mid-infrared from the beginning of December 2011 to beginning of January 2012. Two 5.2' x 5.2' regions near the center of the cluster were observed twice each for about one day with {\it Spitzer}'s high precision staring mode ($<$ 1\%) with a cadence of 0.1 or 1 minute \citep{cod13}. Data for the remaining targets were taken in mapping mode (1 -- 3\% precision) with a cadence of 100 minutes \citep{cod13}.

HD 261903 is the only B type star that was observed in one of the two regions near the cluster center in {\it Spitzer} staring mode using the IRAC2 band at 4.5$\mu$m. The total time base of the IRAC2 light curve is 6.01 days where observations were taken in four parts of one day length each with two gaps of one day length each in between the observations.
Hence, the resolution of the {\it Spitzer} light curve is sufficiently high to study variability on time scales of a few hours. 

The IDL package Cluster Grinder \citep{gut09} was used to generate the light curves from the basic calibrated data (BCD) images released by the Spitzer Science Center (SSC) pipeline (version S19.1). Cluster Grinder provides mosaics for each Astronomical Observation Request (AOR), point source locations and photometric measurements from the mosaics. 
From both the BCDs and the mosaics, aperture photometry was computed using an aperture radius of 2.4'' (2 pixels), and a sky annulus from 2.4'' to 8.4'' (2-6 pixels). A detailed description of the {\it Spitzer} data reduction can be found in \citet{cod14}.

\subsection{Frequency analysis}
\label{frequencyanalysis}

For the frequency analyses, we used the software package Period04 \citep{len05} that combines Fourier and least-squares algorithms. Frequencies were then prewhitened and considered to be significant if their amplitudes exceeded four times the local noise level in the amplitude spectrum \citep{bre93,kus97}.

We verified the analysis using the SigSpec software \citep{ree07}. SigSpec computes significance levels for amplitude spectra of time series with arbitrary time sampling. The probability density function of a given amplitude level is solved analytically and the solution includes dependences on the frequency and phase of the signal.

\subsubsection{HD 47469} \label{star1}

\begin{figure}[htb]
\centering
\includegraphics[width=0.5\textwidth,clip]{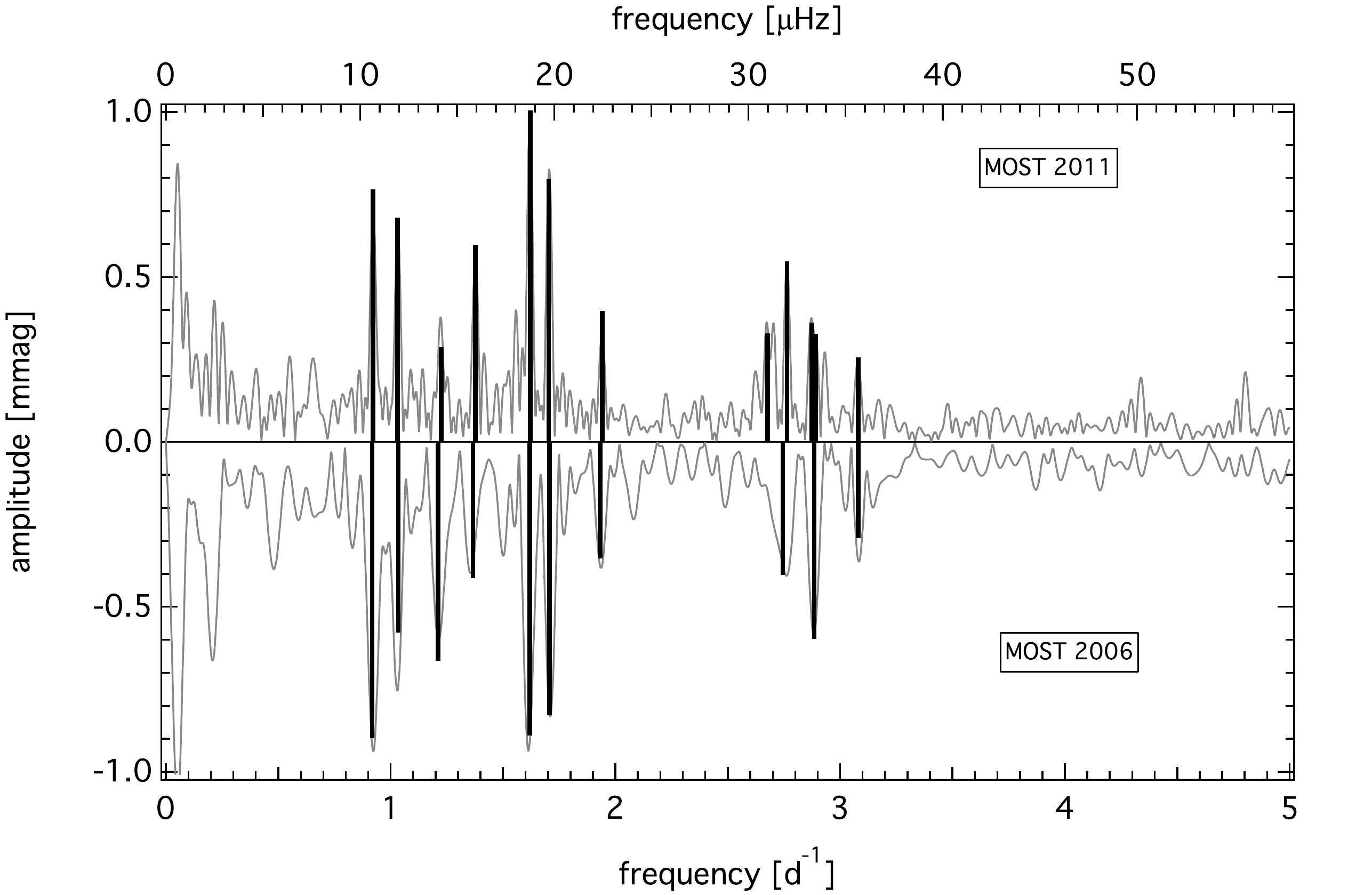}
\caption{Amplitude spectra for HD 47469 obtained from MOST data 2006 (pointing downwards) and 2011/12 (pointing upwards). The identified pulsation frequencies are marked with thicker solid lines.}
\label{hd47469ampspec}
\end{figure}

HD 47469 was observed by the MOST space telescope in 2006 and 2011/12 (Table \ref{obs}) and found to be an SPB type star \citep{zwi09}.
Using both data sets, we could identify 12 pulsation frequencies (see Fig. \ref{hd47469ampspec} and Table \ref{hd47469freqs}). Ten of those were found in both data sets and two additional frequencies could be identified due to the higher frequency resolution only in the 2011/12 data set. We searched for linear combinations and identified F6 at 2.764\cd\,\, to be three times F3 at 0.922\cd. Apart from this, all other frequencies are independent from each other.

Frequency F11 at 1.226\cd\,\, decreased its amplitude from 2006 to 2011/12 significantly by about 0.38 mmag. The amplitude of frequency F4 also decreased between the two observing runs with MOST, but this difference is likely a result of the different frequency resolution: in 2006 the two close frequencies F8 at 2.873 \cd\,\, and F10 2.892 \cd\,\, could not be resolved, hence the corresponding frequency of 2.892\cd\,\, (F10 in 2006) includes the power of the two peaks that were identified separately in the 2011/12 data set. The amplitudes of F1 at 1.621\cd, F5 at 1.377\cd, and F6 at 2.764\cd increased significantly between the 2006 and the 2011/12 observations. Amplitude variability has been reported previously, e.g., for $\delta$ Scuti type stars \citep[e.g.,][]{bow14}. With the present data sets we can only report about the observed change in amplitude for F1, F5 and F6, but cannot conduct a detailed study of their temporal behavior due to the relatively short MOST data sets available to us. Longer photometric time series are needed for a more thorough interpretation of HD 47469's amplitude variability.

\begin{table*}[htb]
\caption{Results from the frequency analysis of HD 47469.} 
\label{hd47469freqs}
\begin{center}
\begin{scriptsize}
\begin{tabular}{lrrrcrrlrrrr}
\hline
\hline
\multicolumn{1}{l}{F} & \multicolumn{2}{c}{frequency} & \multicolumn{1}{c}{P$_1$} 
& \multicolumn{1}{c}{amp$_1$} & \multicolumn{1}{c}{S/N$_1$} & \multicolumn{1}{c}{sig$_1$}  & \multicolumn{1}{c}{amp$_2$} & \multicolumn{1}{c}{S/N$_2$} & \multicolumn{1}{c}{sig$_2$} 
& \multicolumn{1}{c}{amp$_2$ - amp$_1$ }& \multicolumn{1}{c}{lin. combi.} \\
\multicolumn{1}{l}{\#} & \multicolumn{1}{c}{[d$^{-1}$]}  & \multicolumn{1}{c}{[$\mu$Hz]}  & \multicolumn{1}{c}{[d]}
& \multicolumn{1}{c}{[mmag]}  & \multicolumn{1}{c}{ }  & \multicolumn{1}{c}{ }  & \multicolumn{1}{c}{[mmag]}  & \multicolumn{1}{c}{ }  
& \multicolumn{1}{c}{ } & \multicolumn{1}{c}{[mmag]}  & \multicolumn{1}{c}{ }\\
\hline
F1	&	1.621(1)		&	18.76(2)		&	0.6169(5)		&	0.890	&	4.2	&	147.3	&	1.006	&	6.5	&	345.5	&	0.116	&		\\
F2	&	1.703(2)		&	19.71(2)		&	0.5873(6)		&	0.828	&	5.3	&	105.4	&	0.798	&	6.3	&	250.0	&	-0.030	&		\\
F3	&	0.922(2)		&	10.67(2)		&	1.085(2)		&	0.898	&	4.4	&	144.3	&	0.766	&	5.7	&	251.1	&	-0.132	&		\\
F4	&	1.031(2)		&	11.94(2)		&	0.970(1)		&	0.578	&	3.8	&	58.3	&	0.681	&	5.8	&	231.6	&	0.102	&		\\
F5	&	1.377(2)		&	15.94(2)		&	0.726(1)		&	0.413	&	3.7	&	30.4	&	0.598	&	6.7	&	179.7	&	0.185	&		\\
F6	&	2.764(2)		&	31.99(2)		&	0.3618(2)		&	0.403	&	4.4	&	31.3	&	0.548	&	7.3	&	179.5	&	0.145	&	3 * F3	\\
F7	&	1.942(3)		&	22.47(3)		&	0.5150(7)		&	0.353	&	3.0	&	21.1	&	0.397	&	4.6	&	83.6	&	0.044	&		\\
F8	&	2.873(3)		&	33.26(3)		&	0.3480(3)		&	-	&	-	&	-	&	0.361	&	5.6	&	80.9	&	-	&		\\
F9	&	2.677(3)		&	30.98(4)		&	0.3736(4)		&	-	&	-	&	-	&	0.329	&	5.7	&	71.6	&	-	&		\\
F10	&	2.892(3)		&	33.47(4)		&	0.3458(4)		&	0.598	&	5.7	&	59.8	&	0.328	&	5.9	&	55.8	&	-0.271	&		\\
F11	&	1.226(3)		&	14.19(4)		&	0.816(2)		&	0.664	&	3.9	&	69.8	&	0.287	&	3.9	&	57.3	&	-0.377	&		\\
F12	&	3.081(4)		&	35.66(4)		&	0.3245(4)		&	0.292	&	4.3	&	16.4	&	0.257	&	5.1	&	46.5	&	-0.035	&		\\
\hline
\end{tabular}
\end{scriptsize}
\end{center}
\tablefoot{Pulsation frequencies, periods, amplitudes, signal-to-noise values and SigSpec significances identified from the MOST data sets from 2006 (amp$_1$, S/N$_1$,sig$_1$) and 2011/12 (amp$_2$, S/N$_2$, sig$_2$) as well as the the corresponding amplitude differences (amp$_2$ - amp$_1$). The last-digit errors of the frequencies and periods are given in parentheses; frequency errors were computed according to \citet{kal08} and the errors in periods were derived using standard error propagation. Possible linear combinations are given in the last column (lin. combi).}
\end{table*}

\subsubsection{HD 48012} \label{star2}
\begin{figure}[htb]
\centering
\includegraphics[width=0.48\textwidth,clip]{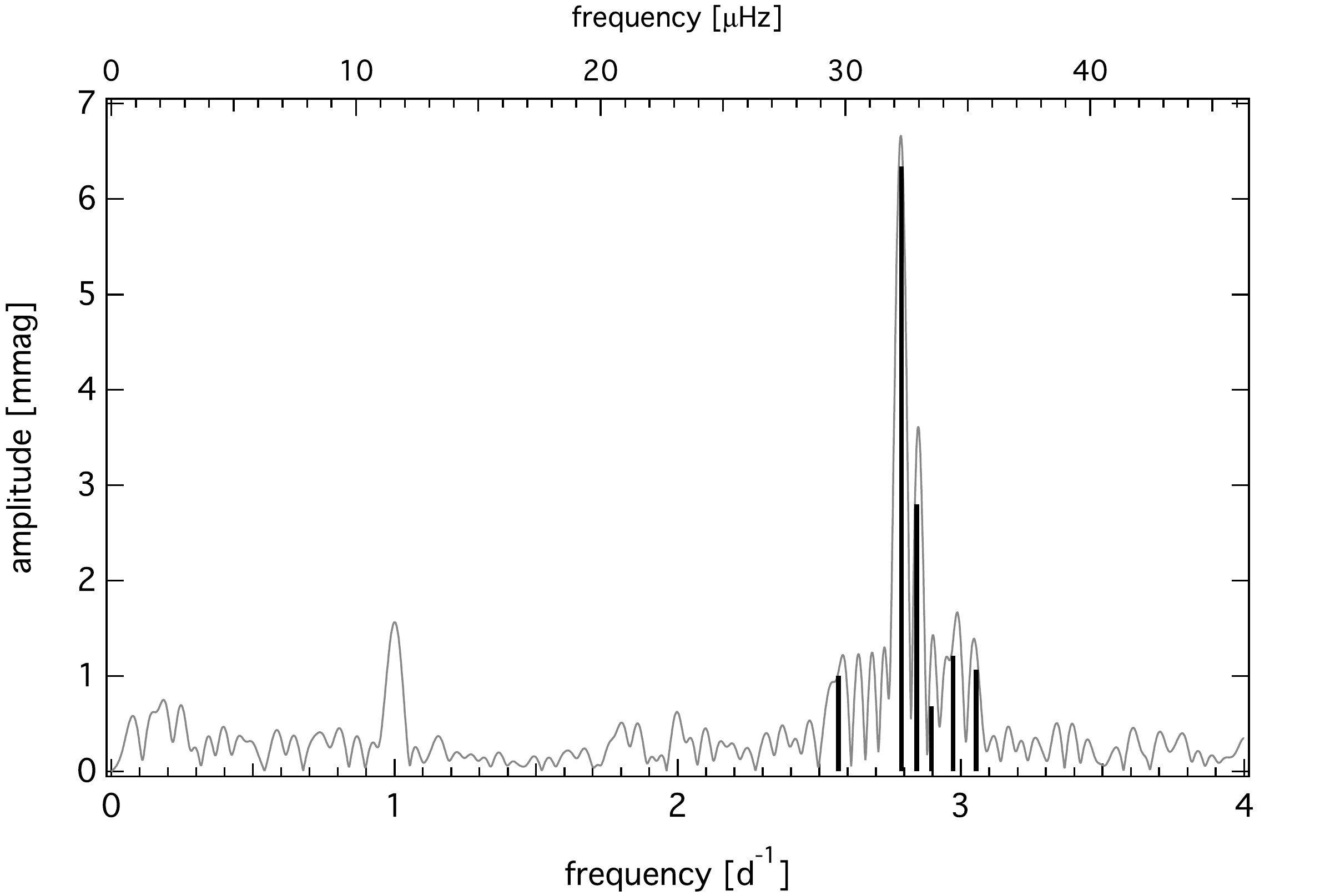}
\caption{Amplitude spectrum for HD 48012 (\#2) obtained from MOST data 2006. The identified pulsation frequencies are marked with black solid lines.}
\label{hd48012ampspec}
\end{figure}

MOST took observations of HD 48012 only in 2006 (Table \ref{obs}).
Figure \ref{hd48012ampspec} shows the original amplitude spectrum of HD 48012 from the 2006 observations with the six identified pulsation frequencies (black solid lines) which are indicative for SPB type pulsation (see Table \ref{hd48012freqs}). It can clearly be seen that additional frequencies in the range between $\sim$2.5 and 3.0d$^{-1}$ are buried in the noise.

\begin{table}[htb]
\caption{Results from the frequency analysis of HD 48012.} 
\label{hd48012freqs}
\begin{center}
\begin{scriptsize}
\begin{tabular}{lrrrcrr}
\hline
\hline
\multicolumn{1}{l}{F} & \multicolumn{2}{c}{frequency} & \multicolumn{1}{c}{P$_1$} 
& \multicolumn{1}{c}{amp$_1$} & \multicolumn{1}{c}{S/N$_1$} & \multicolumn{1}{c}{sig$_1$}   \\
\multicolumn{1}{l}{\#} & \multicolumn{1}{c}{[d$^{-1}$]}  & \multicolumn{1}{c}{[$\mu$Hz]}  & \multicolumn{1}{c}{[d]}
& \multicolumn{1}{c}{[mmag]}  & \multicolumn{1}{c}{ }  & \multicolumn{1}{c}{ }   \\
\hline
F1	&	2.790(2)		&	32.30(2)		&	0.3584(2)		&	6.339	&	10.6	&	524.4	\\
F2	&	2.844(4)		&	32.92(5)		&	0.3516(5)		&	2.802	&	12.5	&	126.5	\\
F3	&	3.054(8)		&	35.35(9)		&	0.3275(9)		&	1.069	&	6.5	&	29.2	\\
F4	&	2.972(9)		&	34.40(10)		&	0.3365(10)	&	1.213	&	6.8	&	25.0	\\
F5	&	2.567(10)		&	29.71(11)		&	0.3896(15)	&	1.005	&	7.5	&	20.7	\\
F6	&	2.896(14)		&	33.52(16)		&	0.3453(17)	&	0.684	&	5.6	&	9.9	\\
\hline
\end{tabular}
\end{scriptsize}
\end{center}
\tablefoot{Pulsation frequencies, periods, amplitudes, signal-to-noise values and SigSpec significances identified from the MOST data set from 2006 (amp$_1$, S/N$_1$,sig$_1$) The last-digit errors of the frequencies and periods are given in parentheses; frequency errors were computed according to \citet{kal08} and the errors in periods were derived using standard error propagation.}
\end{table}


\subsubsection{HD261810} \label{star3}
\begin{figure}[htb]
\centering
\includegraphics[width=0.48\textwidth,clip]{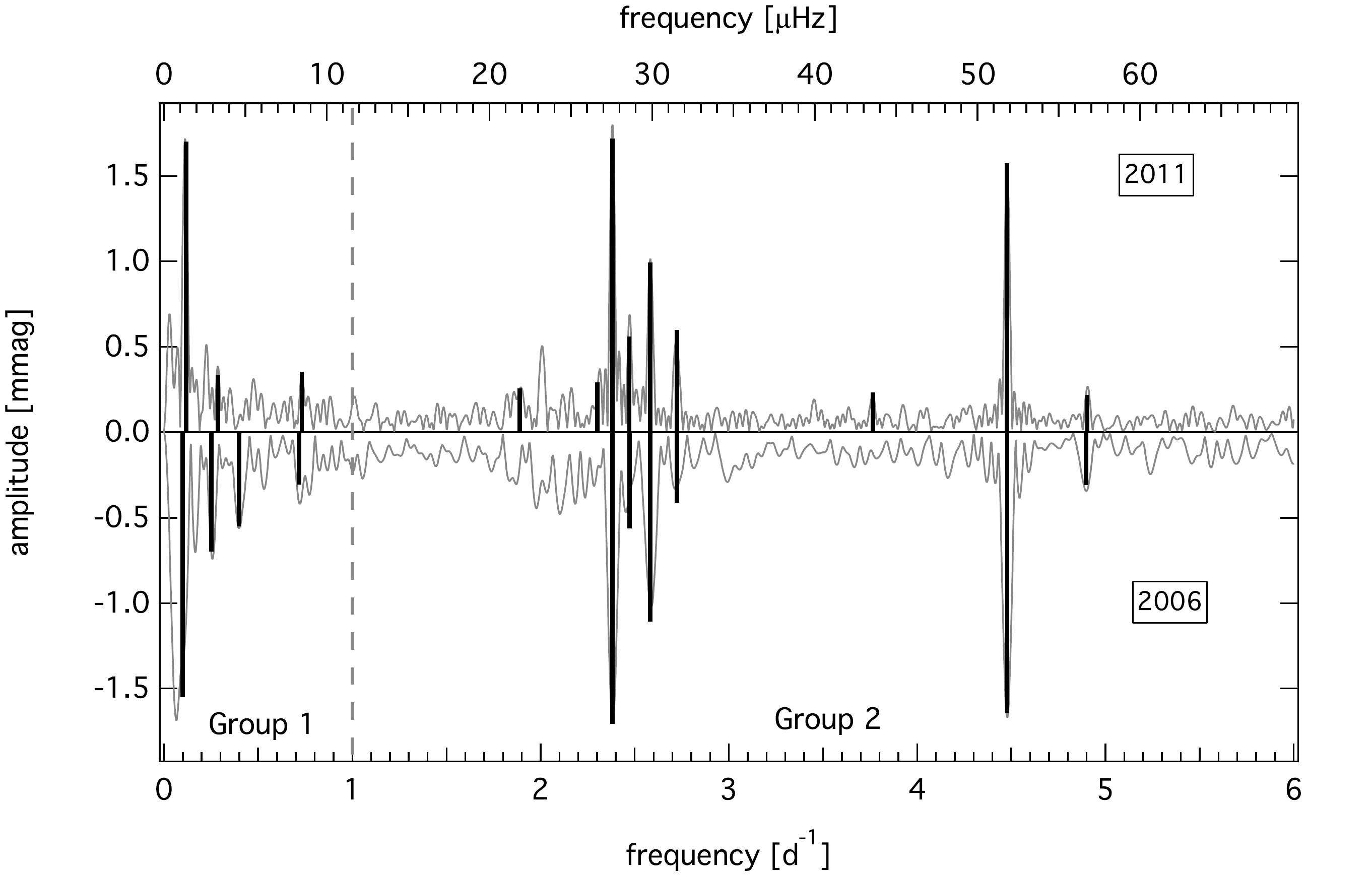}
\caption{Amplitude spectra for HD 261810 obtained from MOST data 2006 (pointing downwards) and 2011/12 (pointing upwards). The identified intrinsic frequencies are marked with black solid lines. Group 1 frequencies are smaller than 1\cd; Group 2 frequencies higher than about 1.8\cd.}
\label{hd261810ampspec}
\end{figure}

HD 261810 was observed by the MOST space telescope in 2006 and in 2011/12 (Table \ref{obs}) and identified as SPB pulsator \citep{zwi09}. 
Our frequency analysis revealed 13 intrinsic frequencies that can be divided in two groups: the lowest frequency range - hereafter named Group 1 - lower than 1\cd, and the higher frequency range - hereafter named Group 2 - from about 1.8 to 5 \cd. Figure \ref{hd261810ampspec} illustrates the original amplitude spectra for the 2006 data (pointing downwards) and the 2011/12 data (pointing upwards). The identified intrinsic frequencies are marked with black solid lines.

Frequencies in Group 1 can all be explained as linear combinations of higher frequencies (see Table \ref{hd261810freqs}). We clearly identify frequencies in Group 2 to be caused by SPB type pulsations.

\begin{table*}[htb]
\caption{Results from the frequency analysis of HD 261810.} 
\label{hd261810freqs}
\begin{center}
\begin{scriptsize}
\begin{tabular}{lrrrcrrlrrrr}
\hline
\hline
\multicolumn{1}{l}{F} & \multicolumn{2}{c}{frequency} & \multicolumn{1}{c}{P$_1$} 
& \multicolumn{1}{c}{amp$_1$} & \multicolumn{1}{c}{S/N$_1$} & \multicolumn{1}{c}{sig$_1$}  & \multicolumn{1}{c}{amp$_2$} & \multicolumn{1}{c}{S/N$_2$} & \multicolumn{1}{c}{sig$_2$} 
& \multicolumn{1}{c}{amp$_2$ - amp$_1$ }& \multicolumn{1}{c}{lin. combi.} \\
\multicolumn{1}{l}{\#} & \multicolumn{1}{c}{[d$^{-1}$]}  & \multicolumn{1}{c}{[$\mu$Hz]}  & \multicolumn{1}{c}{[d]}
& \multicolumn{1}{c}{[mmag]}  & \multicolumn{1}{c}{ }  & \multicolumn{1}{c}{ }  & \multicolumn{1}{c}{[mmag]}  & \multicolumn{1}{c}{ }  
& \multicolumn{1}{c}{ } & \multicolumn{1}{c}{[mmag]}  & \multicolumn{1}{c}{ }\\
\hline
F1	&	2.382(1)	&	27.57(1)	&	0.4198(2)	&	1.707	&	8.8	&	187.0	&	1.720	&	11.0	&	504.0	&	0.013 &	\\
F2	&	0.117(1)	&	1.36(1)	&	8.53(8)	&	-	&	-	&	-	&	1.706	&	10.3	&	528.3	&	- &  F4 - F6	\\
F3	&	4.478(1)	&	51.82(1)	&	0.22334(6)	&	1.643	&	17.6	&	188.2	&	1.574	&	28.6	&	499.3	&	-0.069 & F4 + F11	\\
F4	&	2.583(2)	&	29.90(2)	&	0.3871(2)	&	1.108	&	6.6	&	74.2	&	0.993	&	9.9	&	232.4	&	-0.115 & 	\\
F5	&	2.723(3)	&	31.52(3)	&	0.3672(3)	&	0.414	&	3.5	&	13.3	&	0.597	&	7.3	&	105.3	&	0.183 &	\\
F6	&	2.472(3)	&	28.61(3)	&	0.4045(5)	&	0.561	&	4.2	&	21.8	&	0.560	&	6.9	&	79.3	&	-0.001 &	\\
F7	&	0.43(1)	&	4.9(1)	&	2.34(6)	&	0.432	&	3.5	&	14.0	&	-	&	-	&	-	& - & 	F5 - F10	\\
F8	&	0.732(4)	&	8.47(5)	&	1.366(8)	&	-	&	-	&	-	&	0.354	&	3.6	&	37.2	&	- & F7 + F9	\\
F9	&	0.286(5)	&	3.31(5)	&	3.50(5)	&	0.564	&	3.9	&	20.8	&	0.336	&	3.1	&	32.8	& - 0.228 &	F4 - F10	\\
F10	&	2.302(5)	&	26.64(6)	&	0.434(1)	&	-	&	-	&	-	&	0.294	&	4.0	&	24.4	&	- & 	\\
F11	&	1.889(6)	&	21.86(7)	&	0.530(2)	&	-	&	-	&	-	&	0.255	&	3.6	&	19.1	&	- & 	\\
F12	&	3.766(6)	&	43.59(7)	&	0.2655(5)	&	-	&	-	&	-	&	0.233	&	4.6	&	16.0	&	- &  2 * F11	\\
F13	&	4.904(7)	&	56.76(8)	&	0.2039(3)	&	0.309	&	4.2	&	7.8	&	0.218	&	4.7	&	14.3	&	-0.090 & F4 + F10	\\
\hline
\end{tabular}
\end{scriptsize}
\end{center}
\tablefoot{Same format as in Table \ref{hd47469freqs}.}
\end{table*}

\subsubsection{HD 261878} \label{star4}
\begin{figure}[htb]
\centering
\includegraphics[width=0.48\textwidth,clip]{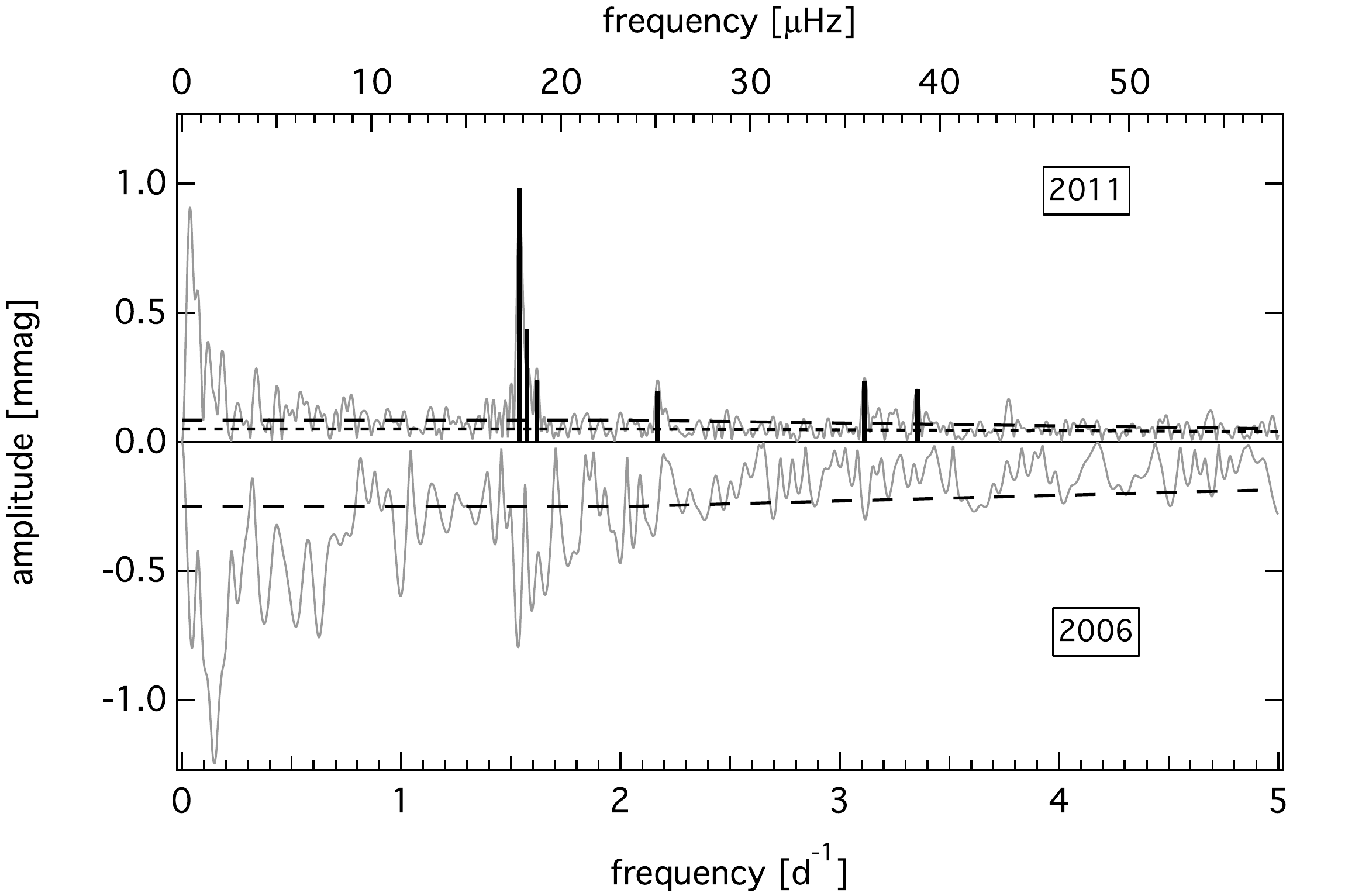}
\caption{Amplitude spectra for HD 261878 obtained from MOST data 2006 (pointing downwards) and 2011/12 (pointing upwards). Dashed lines mark the noise spectrum for the original 2006 and 2011/12 data, respectively; the dotted line shows the noise spectrum of the residuals of the 2011/12 data set after prewhitening all significant frequencies.}
\label{hd261878ampspec}
\end{figure}

Observations for HD 261878 were taken in 2006 and in 2011/12 by the MOST satellite (Table \ref{obs}).
The frequency analysis of the 2006 data set revealed no formally significant frequency while the 2011/12 data set shows six frequencies that can be attributed to SPB type pulsation. The reason for this difference between the two data sets is the rather high noise level in the 2006 data (see Fig. \ref{hd261878ampspec}). 
The average noise level of the original 2006 data set was calculated to be 137 ppm in the range from 0 to 100\cd\,\, while for the 2011/12 data set, the same average noise lies at 39 ppm. A comparison of the amplitude spectra for both MOST observing runs shows that the frequencies we identified from the 2011/12 data sets are buried in the noise of the 2006 data set.


\begin{table}[htb]
\caption{Results from the frequency analysis of HD 261878.} 
\label{hd261878freqs}
\begin{center}
\begin{scriptsize}
\begin{tabular}{lrrrcrr}
\hline
\hline
\multicolumn{1}{l}{F} & \multicolumn{2}{c}{frequency} & \multicolumn{1}{c}{P$_2$} 
& \multicolumn{1}{c}{amp$_2$} & \multicolumn{1}{c}{S/N$_2$} & \multicolumn{1}{c}{sig$_2$}   \\
\multicolumn{1}{l}{\#} & \multicolumn{1}{c}{[d$^{-1}$]}  & \multicolumn{1}{c}{[$\mu$Hz]}  & \multicolumn{1}{c}{[d]}
& \multicolumn{1}{c}{[mmag]}  & \multicolumn{1}{c}{ }  & \multicolumn{1}{c}{ }   \\
\hline
F1	&	1.539(1)	&	17.82(2)	&	0.6496(6)		&	0.985	&	12.3	&	288.5	\\
F2	&	1.573(3)	&	18.21(3)	&	0.6355(11)	&	0.437	&	7.6	&	87.6	\\
F3	&	1.618(6)	&	18.736)	&	0.6178(21)	&	0.239	&	4.1	&	21.7	\\
F4	&	2.170(6)	&	25.12(7)	&	0.4608(13)	&	0.196	&	4.0	&	18.4	\\
F5	&	3.114(5)	&	36.05(6)	&	0.3211(5)		&	0.234	&	4.7	&	24.6	\\
F6	&	3.354(6)	&	38.82(7)	&	0.2981(5)		&	0.205	&	4.4	&	20.1	\\
\hline
\end{tabular}
\end{scriptsize}
\end{center}
\tablefoot{Pulsation frequencies, periods, amplitudes, signal-to-noise values and SigSpec significances identified from the MOST data set from 2011 (amp$_2$, S/N$_2$,sig$_2$) The last-digit errors of the frequencies and periods are given in parentheses; frequency errors were computed according to \citet{kal08} and the errors in periods were derived using standard error propagation.}
\end{table}



\subsubsection{HD 47777} \label{star5}
The star HD 47777 was discussed in detail by \citet{fos14}. Here we summarize HD 47777's properties for completeness.

The MOST data for HD 47777 obtained in 2006 and 2011/12 revealed a single intrinsic frequency at 0.3787\cd\, corresponding to a rotation period of 2.641 days. 
The non-LTE analysis of high-resolution spectroscopic data for HD 47777 yielded an effective temperature, \Teff, of 22000$\pm$1000\,K, a surface gravity, \logg, of 4.2$\pm$0.1 and a projected rotational velocity, \vsini, of 60\,\kms\, \citep{fos14}. Using spectropolarimetric data obtained with the \espa\, spectropolarimeter of the Canada-France-Hawaii Telescope (CFHT) the authors reported the detection of a magnetic field with a strength of the longitudinal component of 469$\pm$87\,G \citep{fos14}.


\subsubsection{HD 47887} \label{star6}
The MOST space telescope observed HD 47887 only in 2006. A detailed discussion on HD 47887 is given in \citet{fos14}, hence, here we only give a brief summary of the star's properties.

HD 47887 shows a single intrinsic frequency at 0.513\cd\, corresponding to a period of 1.949 days.
From high-resolution spectroscopy, \citet{fos14} found \Teff\, to be 24000$\pm$1000\,K, \logg\, as 4.1$\pm$0.1 and \vsini\, as 45\,\kms. Spectropolarimetric measurements taken with the \espa\, spectropolarimeter of the Canada-France-Hawaii Telescope (CFHT) revealed the presence of a magnetic field with a longitudinal component of 373$\pm$48\,G \citep{fos14}.


\subsubsection{HD 261903} \label{star7}
\begin{figure}[htb]
\centering
\includegraphics[width=0.48\textwidth,clip]{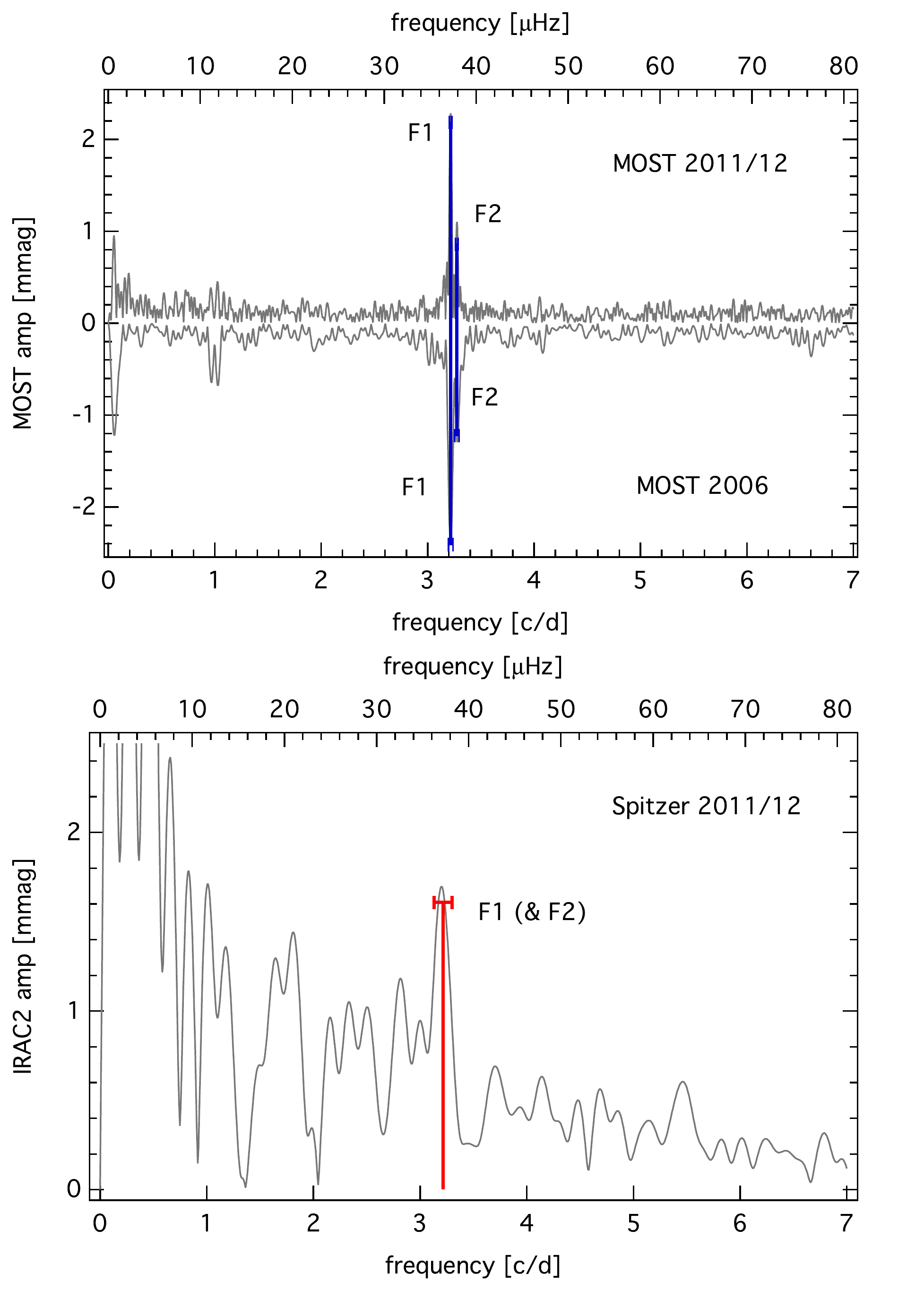}
\caption{Amplitude spectra of HD 261903  from the MOST observations in 2006 (upper panel pointing downwards), in 2011/12 (upper panel pointing upwards) and from the Spitzer observations in 2011/12 (lower panel).}
\label{hd261903amps}
\end{figure}

For HD 261903 we were able to investigate two MOST data sets and an infrared light curve obtained by the Spitzer space telescope (Table \ref{obs}).
Figure \ref{hd261903amps} shows the respective amplitude spectra from the MOST 2006 (upper panel pointing downwards) and 2011/12 (upper panel pointing upwards) data sets where the two frequencies F1 at 3.216\,\cd\ and F2 at 3.275\,\cd\ (see Table \ref{hd261903freqs}) are marked as solid lines. 
Figure \ref{hd261903amps} shows the amplitude spectrum of the IRAC2 light curve (bottom panel). It can clearly be seen that the frequency resolution is not sufficient to resolve both frequencies found in the MOST data. With only two very close frequencies in a rapidly rotating star (see Paper II), we interpret the variability to be caused by differential rotation of a spotted star.

\begin{table*}[htb]
\caption{Results from the frequency analysis of HD 261903 using MOST and {\it Spitzer} data.} 
\label{hd261903freqs}
\begin{center}
\begin{scriptsize}
\begin{tabular}{lrrrcrlrrrrr}
\hline
\hline
\multicolumn{1}{l}{ } & \multicolumn{2}{c}{ } & \multicolumn{3}{c}{MOST2006} & \multicolumn{3}{c}{MOST2011/12} & \multicolumn{3}{c}{{\it Spitzer} IRAC2}  \\
\hline
\multicolumn{1}{l}{F} & \multicolumn{2}{c}{frequency} 
& \multicolumn{1}{c}{amp$_1$} & \multicolumn{1}{c}{phase$_1$} & \multicolumn{1}{c}{S/N$_1$}   & \multicolumn{1}{c}{amp$_2$} & \multicolumn{1}{c}{phase$_2$} & \multicolumn{1}{c}{S/N$_2$} & \multicolumn{1}{c}{amp$_3$} & \multicolumn{1}{c}{phase$_3$}  & \multicolumn{1}{c}{S/N$_3$}   \\
\multicolumn{1}{l}{\#} & \multicolumn{1}{c}{[d$^{-1}$]}  & \multicolumn{1}{c}{[$\mu$Hz]}  & \multicolumn{1}{c}{[mmag]}  & \multicolumn{1}{c}{ }   & \multicolumn{1}{c}{ }  & \multicolumn{1}{c}{[mmag]}  & \multicolumn{1}{c}{ }   & \multicolumn{1}{c}{ }  & \multicolumn{1}{c}{[mmag]}    & \multicolumn{1}{c}{ } & \multicolumn{1}{c}{ }  \\
\hline
F1 & 3.216(2) & 37.22(3) & 2.408 & 0.211 & 9.9 &  2.187 & 0.459 & 19.9 & 1.609 & 0.381 & 4.4 \\
 F2 & 3.275(4) & 37.91(7) &  1.221 & 0.682 & 12.3  &  0.860 & 0.131 & 9.2  & ... & ... & ... \\
\hline
\end{tabular}
\end{scriptsize}
\end{center}
\tablefoot{Intrinsic frequencies, amplitudes and signal-to-noise values identified from the MOST data sets from 2006 (amp$_1$, ,phase$_1$ S/N$_1$) and 2011/12 (amp$_2$, phase$_2$, S/N$_2$) and the {\it Spitzer IRAC2} data sets from 2011/12 (amp$_3$,  phase$_3$, S/N$_3$). Frequency errors were computed according to \citet{kal08}.}
\end{table*}


\subsubsection{HD 261938} \label{star8}
\begin{figure}[htb]
\centering
\includegraphics[width=0.48\textwidth,clip]{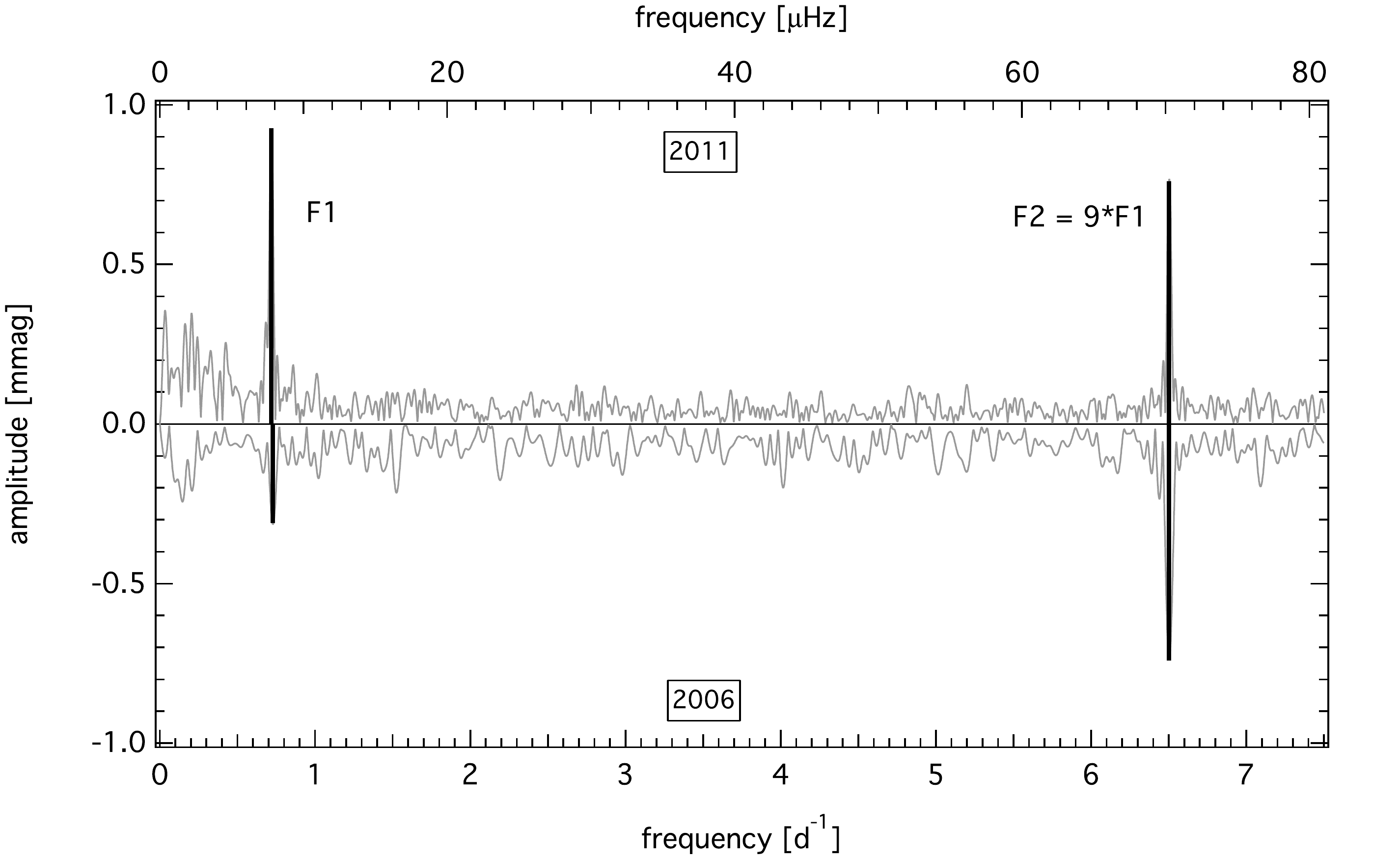}
\caption{Amplitude spectra for HD 261938 obtained from MOST data 2006 (pointing downwards) and 2011/12 (pointing upwards). }
\label{hd261938ampspec}
\end{figure}

The analysis of the two MOST data sets for HD 261938 (see Table \ref{obs}) revealed two significant frequencies: F1 at 0.71845\cd, and F2 at 6.50342\cd\,\, (see Fig. \ref{hd261938ampspec}). F1 corresponds to a period, P1 of about 1.392\cd, which could be caused by pulsations, binarity or rotational modulation of a spotted surface. 
F2 on the other hand is much too high to be explained as, e.g., SPB type pulsation. But a closer inspection revealed that F2 is nine times F1 within the given errors. 
With the present observational material we cannot decide whether F2 is indeed the ninth harmonic of F1 and the "missing'' harmonics are buried in the noise or whether F2 has a different origin. Hence, we cannot draw a final conclusion of HD 261938's variability. We can only speculate that we might either see rotational modulation of a spotted surface or that we might observe a binary system with tidally induced pulsations.

\begin{table*}[htb]
\caption{Results from the frequency analysis of HD 261938.} 
\label{hd261938freqs}
\begin{center}
\begin{scriptsize}
\begin{tabular}{lrrrcrrcrr}
\hline
\hline
\multicolumn{1}{l}{F} & \multicolumn{2}{c}{frequency} & \multicolumn{1}{c}{P$_1$} 
& \multicolumn{1}{c}{amp$_1$} & \multicolumn{1}{c}{S/N$_1$} & \multicolumn{1}{c}{sig$_1$} & \multicolumn{1}{c}{amp$_2$} & \multicolumn{1}{c}{S/N$_2$} & \multicolumn{1}{c}{sig$_2$}   \\
\multicolumn{1}{l}{\#} & \multicolumn{1}{c}{[d$^{-1}$]}  & \multicolumn{1}{c}{[$\mu$Hz]}  & \multicolumn{1}{c}{[d]}
& \multicolumn{1}{c}{[mmag]}  & \multicolumn{1}{c}{ }  & \multicolumn{1}{c}{ }  & \multicolumn{1}{c}{[mmag]}  & \multicolumn{1}{c}{ }  & \multicolumn{1}{c}{ }\\
\hline
F1	&	0.719(1)		&	8.32(2)		&	1.392(3)		&	0.316	&	3.7	&	10.85	& 0.927	& 7.54	& 388.6	\\
F2	&	6.503(2)		&	75.27(2)		&	0.154(1)		&	0.741	&	3.7	&	55.77	& 0.760	& 19.6	& 289.7	\\
\hline
\end{tabular}
\end{scriptsize}
\end{center}
\tablefoot{Pulsation frequencies, periods, amplitudes, signal-to-noise values and SigSpec significances identified from the MOST data set from 2006 (amp$_1$, S/N$_1$,sig$_1$) and 2011/12 (amp$_2$, S/N$_2$,sig$_2$). The last-digit errors of the frequencies and periods are given in parentheses; frequency errors were computed according to \cite{kal08} and the errors in periods were derived using standard error propagation.}
\end{table*}


\subsubsection{NGC 2264 137} \label{star9}
NGC 2264 137 was observed by MOST in 2006 and 2011/12; additionally it was also included as imagette in the CoRoT Short Run SRa01 in 2008. 

\begin{figure}[htb]
\centering
\includegraphics[width=0.48\textwidth,clip]{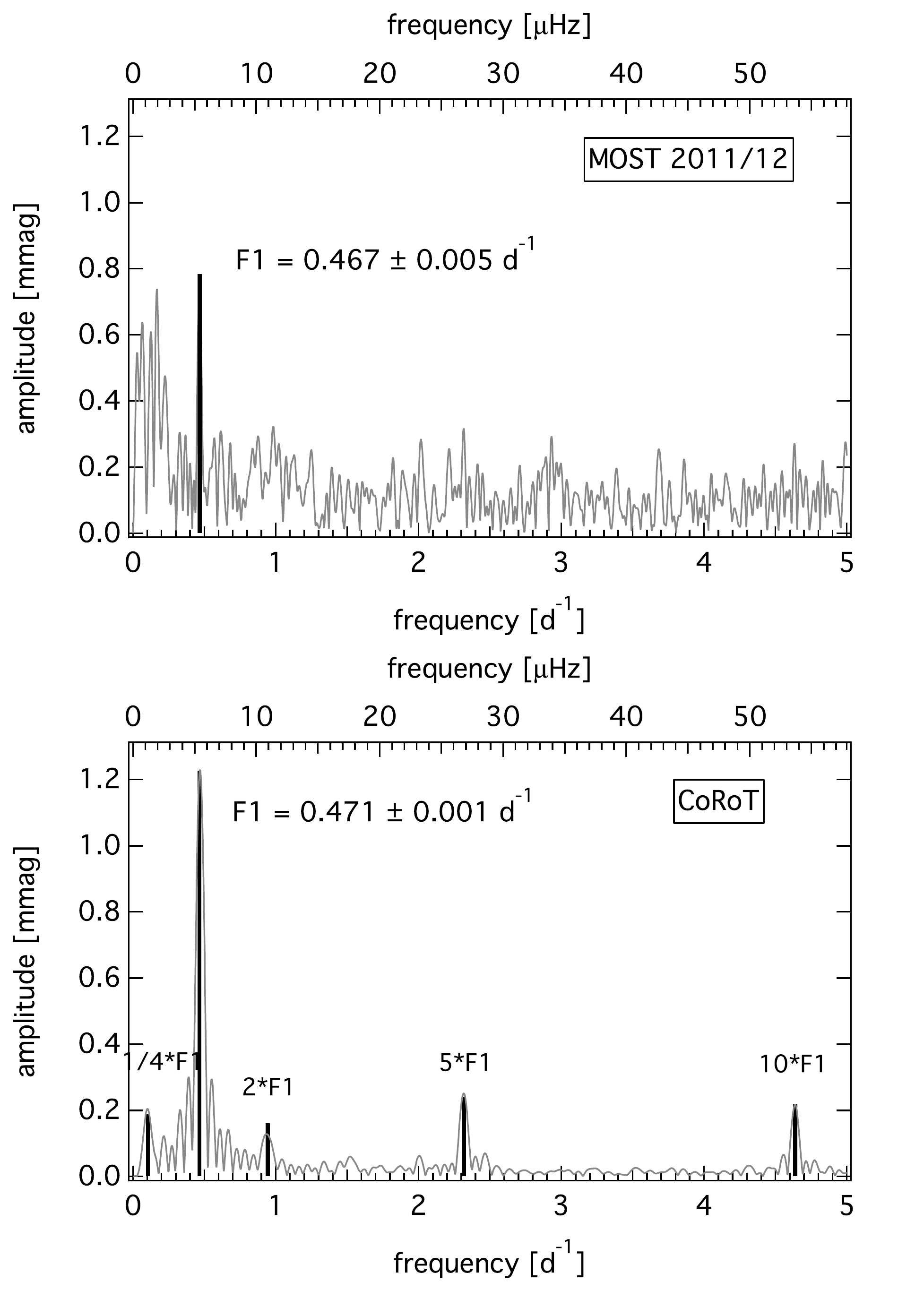}
\caption{Amplitude spectrum of NGC 2264 137 from the MOST observations in 2011/12 (top panel) and the CoRoT SRa01 observations in 2008 (bottom panel).}
\label{ngc2264w137amps}
\end{figure}

For NGC 2264 137 we identified five formally significant frequencies from the CoRoT data set where four can be explained as multiples of the frequency with the highest amplitude (F1) at 0.471\cd\,\, as illustrated in the bottom panel of Fig. \ref{ngc2264w137amps}. F1 also appears in the MOST data set obtained in 2011/12 (top panel in Fig. \ref{ngc2264w137amps}). The MOST 2006 data set unfortunately has a significantly higher noise level which makes it impossible to identify a frequency with an amplitude at the millimagnitude level. 
Table \ref{ngc2264w137freqs} lists the significant frequencies for NGC 2264 137 which we attribute to be caused by rotational modulation of a spotted surface. The rotational period would then be the inverse of F1, i.e., 2.123 days.

\begin{table*}[htb]
\caption{Results from the frequency analysis of NGC 2264 137 using CoRoT and MOST data.} 
\label{ngc2264w137freqs}
\begin{center}
\begin{tabular}{lrrrrrrrrr}
\hline
\hline
\multicolumn{1}{l}{F} & \multicolumn{2}{c}{frequency}  
& \multicolumn{1}{c}{amp$_C$} & \multicolumn{1}{c}{S/N$_C$} & \multicolumn{1}{c}{sig$_C$}  & \multicolumn{1}{c}{amp$_2$} & \multicolumn{1}{c}{S/N$_2$} & \multicolumn{1}{c}{sig$_2$}& \multicolumn{1}{c}{multiples}    \\
\multicolumn{1}{l}{\#} & \multicolumn{1}{c}{[d$^{-1}$]}  & \multicolumn{1}{c}{[$\mu$Hz]}  
& \multicolumn{1}{c}{[mmag]}  & \multicolumn{1}{c}{ }  & \multicolumn{1}{c}{ } & \multicolumn{1}{c}{[mmag]} & \multicolumn{1}{c}{ } & \multicolumn{1}{c}{ }  & \multicolumn{1}{c}{ }   \\
\hline
F1	&	0.471(1)		&	5.46(1)			&	1.227	&	6.9	&	3236.2	&	0.784 & 3.8	&	23.0 & \\
F2	&	2.317(4)		&	26.82(5)			&	0.241	&	14.3	&	201.5	& -	& -	& -	& 5 * F1		\\
F3	&	4.638(4)		&	53.68(5)				&	0.218	&	20.0	&	169.2 & -	& -	& -	&	10 * F1	\\
F4	&	0.944(6)		&	10.93(7)			&	0.161	&	6.3	&	93.5	& -	& -	& -	& 2 * F1		\\
F5	&	0.126(6)		&	1.46(7)			&	0.168	&	7.2	&	82.1	& -	& - 	& -	& F1 / 4		\\
\hline
\end{tabular}
\end{center}
\tablefoot{Significant frequencies, amplitudes, signal-to-noise values and SigSpec significances identified from the CoRoT data set from 2008 (amp$_C$, S/N$_C$,sig$_C$) and the MOST data set from 2011/12 (amp$_2$, S/N$_2$,sig$_2$), and the identified multiples of F1 (multiples). The last-digit errors of the frequencies and periods are given in parentheses; frequency errors were computed according to \citet{kal08} and the errors in periods were derived using standard error propagation.}
\end{table*}

\subsubsection{HD 261054} \label{star10}
MOST time series photometry was taken for HD 261054 only in 2006 (Table \ref{obs}).
The frequency analysis revealed only three significant intrinsic frequencies (see Table \ref{hd261054freqs}). Although F3 has a value of nearly exactly 2\cd, we consider it intrinsic to the star and not related to the properties of the data set as the spectral window function of the corresponding field B of the MOST 2006 observations does not show a signal at 2\cd\ (see Fig. \ref{hd47469spws}).
The multi-sine fit to the data does not represent the observed variability perfectly, which leads us to suggest that additional frequencies might be buried in the noise and could be revealed with longer time series. With such a low number of detected frequencies the origin of the star's variability might be connected to pulsations or caused by rotational modulation due to spots. 

HD 261054 is very likely not a member of NGC 2264 because it is located furthest away from the cluster center and at a larger distance \citep[see Section \ref{star_description};][]{gaia16}, has an offset position in the cluster CMD compared to the other B type stars and its proper motion values are significantly different from the overall cluster values as discussed earlier. 
As HD 261054 is listed as emission line star in the literature, its variations are likely to be connected to the Be phenomenon \citep[e.g.,][]{riv13}. For a more detailed study of HD 261054 and a possible interpretation as a Be star, longer photometric time series are required.

\begin{figure}[htb]
\centering
\includegraphics[width=0.48\textwidth,clip]{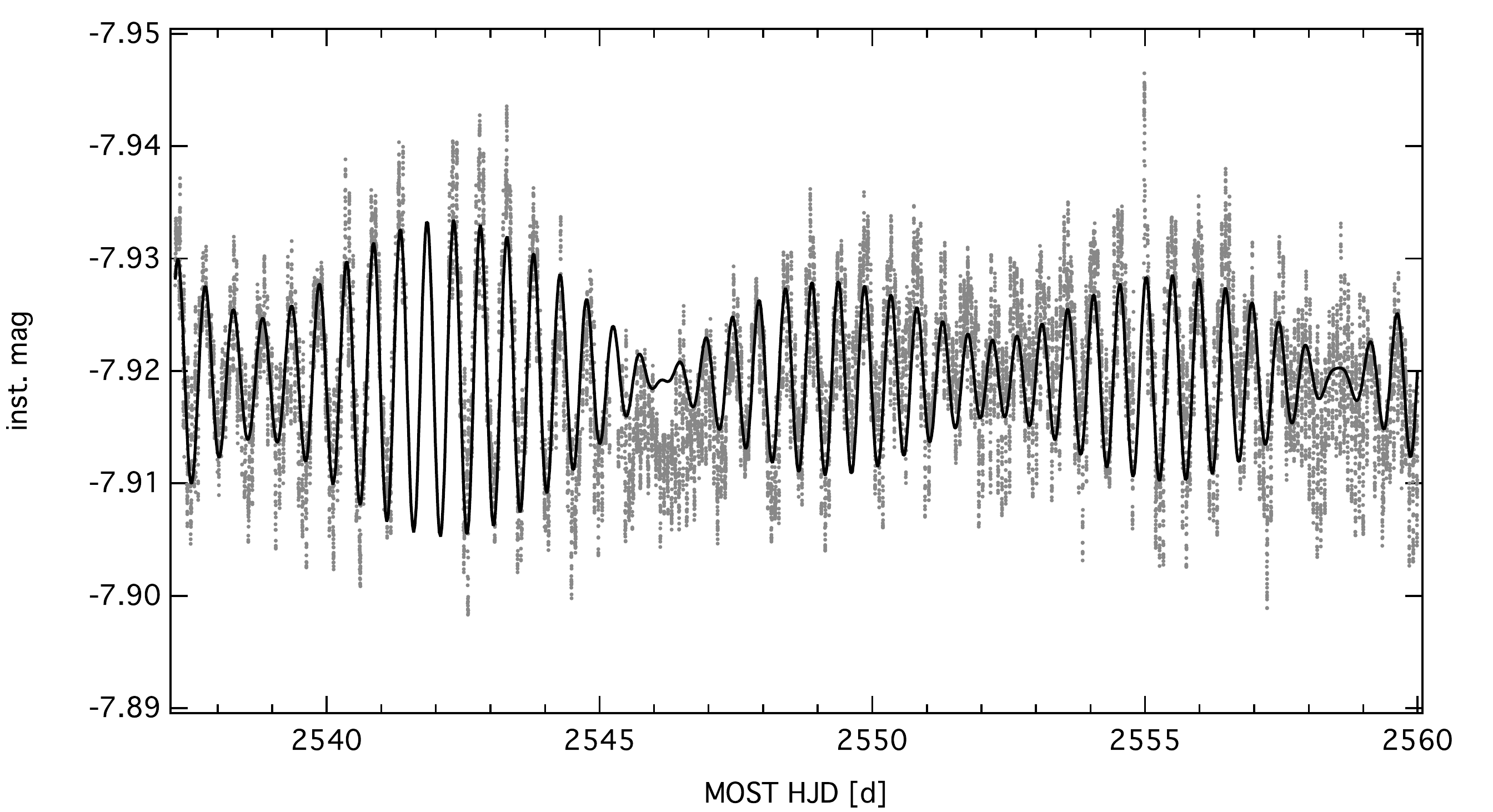}
\caption{Light curve of HD 261054 (grey points) and multi-sine fit with the three intrinsic frequencies (solid black line).}
\label{hd261054lc}
\end{figure}

\begin{figure}[htb]
\centering
\includegraphics[width=0.48\textwidth,clip]{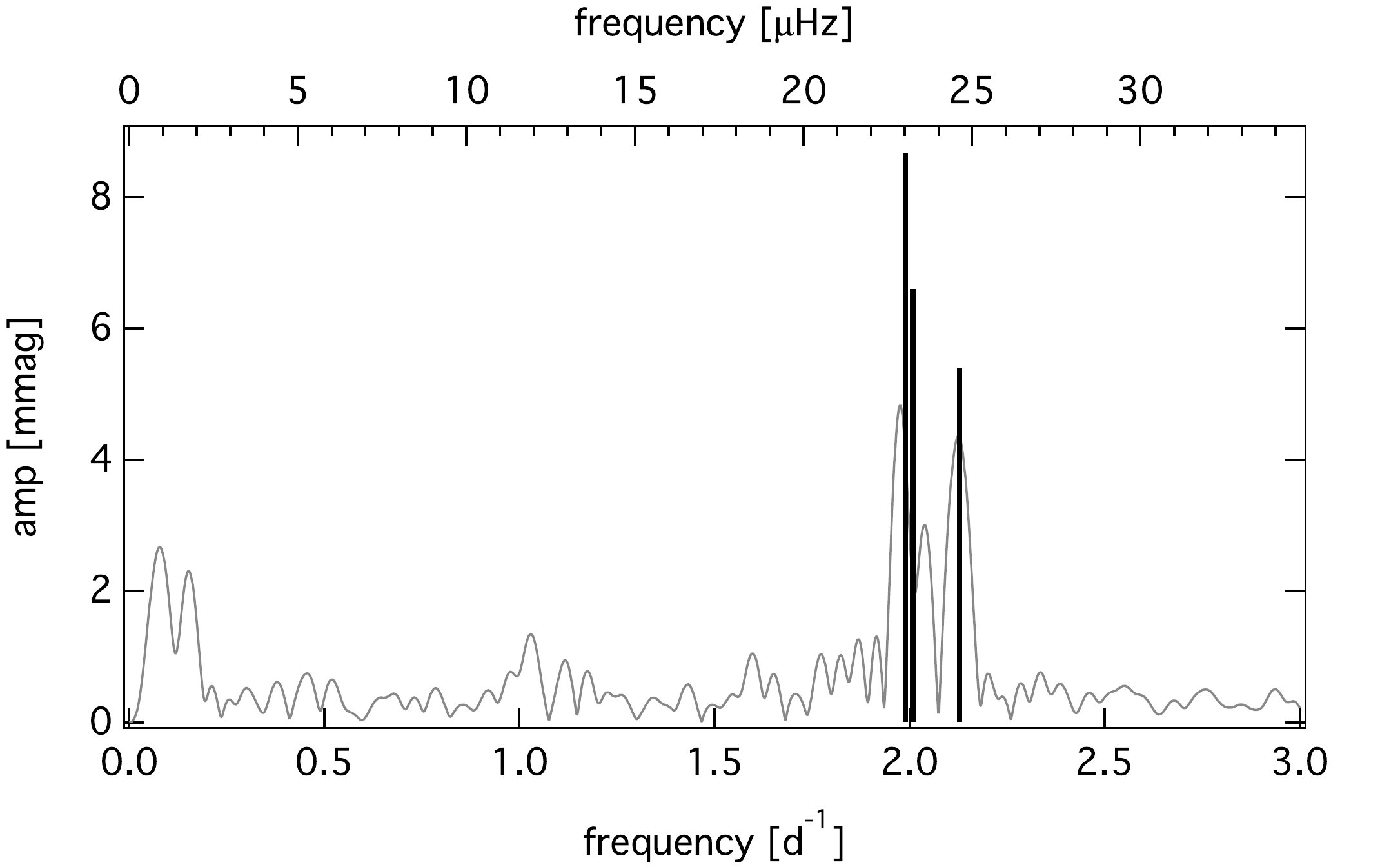}
\caption{Amplitude spectrum of HD 261054 (grey points) obtained from MOST 2006 data. The identified pulsation frequencies are marked with black solid lines.}
\label{hd261054ampspec}
\end{figure}

\begin{table}[htb]
\caption{Results from the frequency analysis of HD 261054.} 
\label{hd261054freqs}
\begin{center}
\begin{scriptsize}
\begin{tabular}{lrrrcrr}
\hline
\hline
\multicolumn{1}{l}{F} & \multicolumn{2}{c}{frequency} & \multicolumn{1}{c}{P$_1$} 
& \multicolumn{1}{c}{amp$_1$} & \multicolumn{1}{c}{S/N$_1$} & \multicolumn{1}{c}{sig$_1$}   \\
\multicolumn{1}{l}{\#} & \multicolumn{1}{c}{[d$^{-1}$]}  & \multicolumn{1}{c}{[$\mu$Hz]}  & \multicolumn{1}{c}{[d]}
& \multicolumn{1}{c}{[mmag]}  & \multicolumn{1}{c}{ }  & \multicolumn{1}{c}{ }   \\
\hline
F1	&	1.989(2)		&	23.02(3)		&	0.5028(6)		&	8.681	&	6.4	&	372.5	\\
F2	&	2.128(2)		&	24.63(2)		&	0.4700(5)		&	5.396	&	15.2	&	427.5	\\
F3	&	2.008(3)		&	23.24(3)		&	0.4981(7)		&	6.597	&	26.4	&	218.5	\\
\hline
\end{tabular}
\end{scriptsize}
\end{center}
\tablefoot{Same format as Table \ref{hd48012freqs}.}
\end{table}


\subsubsection{HD 47961} \label{star11}

\begin{figure}[htb]
\centering
\includegraphics[width=0.48\textwidth,clip]{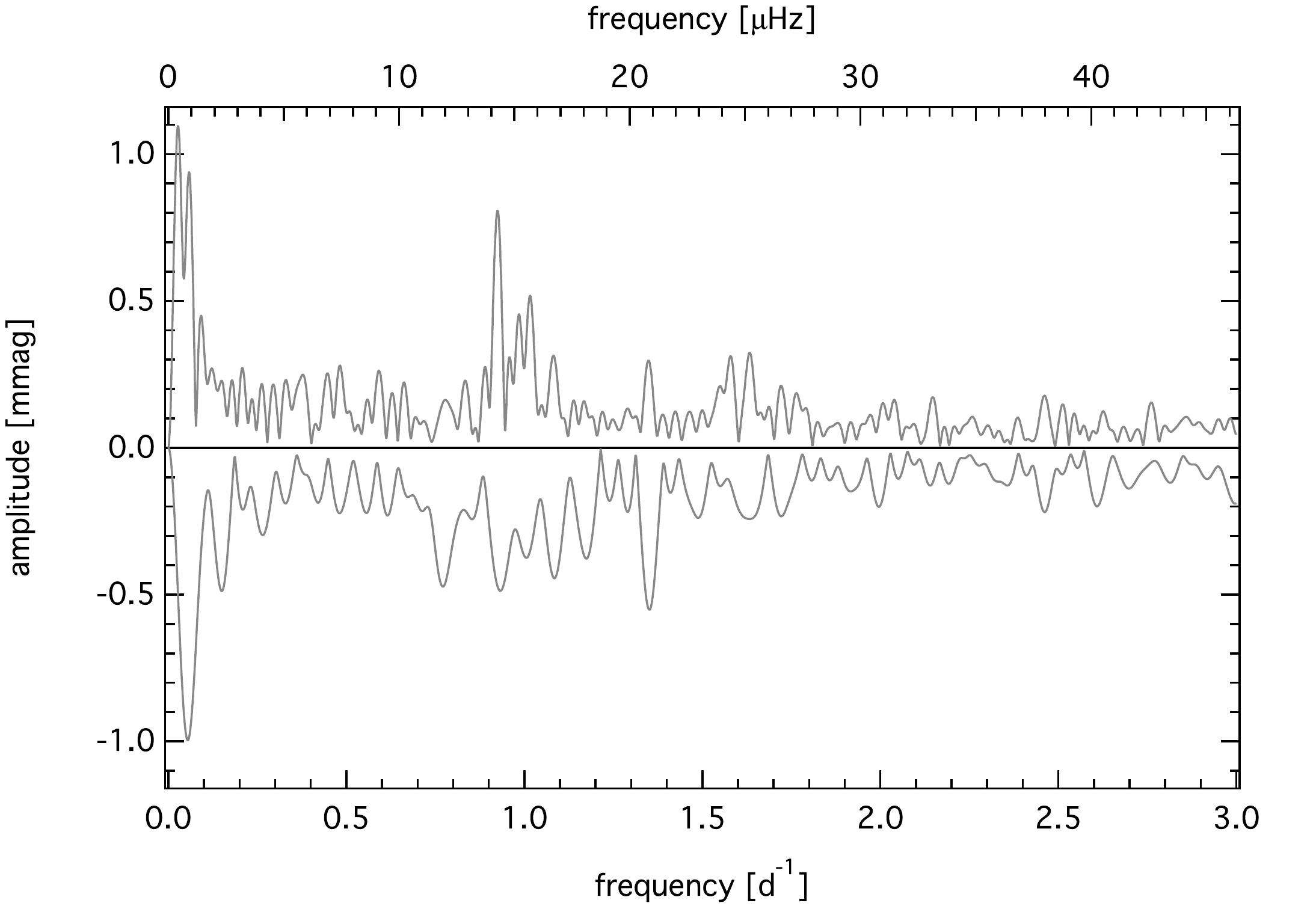}
\caption{Amplitude spectra for HD 47961 obtained from MOST data 2006 (pointing downwards) and 2011/12 (pointing upwards). }
\label{hd47961ampspec}
\end{figure}

MOST observed HD 47961 in 2006 and in both fields in 2011/12. As field 1 and 2 were observed alternatingly in 2011/12, we merged the two data sets and ran the frequency analysis on the complete time series which consists of 25059 data points obtained during a total time base of 39.98 days. In the 2006 data set we find one significant frequency at $\sim$0.055\cd\ and several peaks at similar amplitudes in the region between 0.77 and 1.35\cd\ that are not formally significant (see Fig. \ref{hd47961ampspec}). As the complete 2011/12 data set has a lower noise level and smaller frequency resolution, we can identify the low frequency peak in fact to consist of two frequencies F1 at 0.027\cd\ and F2 at 0.053\cd, where F2 is twice F1, and we find a third frequency F3 at 0.924\cd. The origin of the photometric variability per se is inconclusive. 
Only in combination with the spectroscopic data are we able to suggest the source of the light variations as binarity (see Paper II).

\section{Evolutionary and Asteroseismic Inference}\label{sec-models}


\subsection{Fundamental parameters}

To conduct an evolutionary and asteroseismic inference, it is
crucial to determine the stars' fundamental parameters. Therefore, we
obtained high-resolution, high signal-to-noise spectroscopic and
spectropolarimetric data for all eleven B stars in NGC 2264. A detailed
analysis of all data will be discussed in Paper II. Here we only
describe the determination of the 
atmospheric parameters -- \Teff, \logg, \vsini\ -- for the four SPB type pulsators 
that are used as input for our asteroseismic analysis.

High-resolution spectroscopic data for HD 47469, HD 48012 and HD 261810 were obtained on 2012, January 31 with the \espa\ spectropolarimeter of the Canada-France-Hawaii Telescope (CFHT). \espa\ consists of a table-top, cross-dispersed \'echelle spectrograph fed via a double optical fiber directly from a Cassegrain-mounted polarisation analysis module. Both Stokes $I$ and $V$ spectra were obtained throughout the 3700--10400\,\AA\ spectral range at a resolution of about 65\,000. The spectra were reduced using the Libre-ESpRIT reduction pipeline \citep{don97}. The spectra were then normalized by fitting a low-order polynomial to carefully selected continuum points. 
The signal-to-noise (S/N) ratios of the spectra for HD 47469, HD 48012 
and HD 261810 were 275, 162 and 261 
per 1.8\kms, respectively, measured at a
reference wavelength of 5000\,{\AA}.

The spectrum for HD 261878 was observed on 2011, December 31 with the Robert G. Tull Coud\'e Spectrograph (TS), which is mounted on the 2.7-m telescope of Mc\,Donald Observatory and has a resolving power R $\sim$ 52\,000. The spectra cover a wavelength range of 3633--10849\,\AA\ with gaps between the \'echelle orders at wavelengths longer than 5880\,\AA. 
Bias and flat field frames were obtained at the beginning of each night, while several Th-Ar comparison lamp spectra were obtained each night for wavelength calibration purposes. The reduction was performed using the Image Reduction and Analysis Facility\footnote{IRAF (http://iraf.noao.edu) is distributed by the National Optical Astronomy Observatory, which is operated by the Association of Universities for Research in Astronomy (AURA) under cooperative agreement with the National Science Foundation.} (IRAF). The spectra were normalised by fitting a low order polynomial to carefully selected continuum points. 
As H$\alpha$ is not covered in TS spectra and H$\beta$ is affected by a 
reflection in the imaging system of the spectrograph, we 
employed the H$\gamma$ line in the atmospheric parameter determination in  HD 261878. 
We normalized the H$\gamma$ line continuum using the artificial flat-fielding technique described in \citet{bar02}, which has proven to be successful with TS data \citep[e.g.,][]{zwi13}. The S/N of HD 261878' s spectrum is 101
per pixel at $\lambda \sim$5000{\AA}.

For the determination of the atmospheric parameters, we computed synthetic spectra 
with a hybrid non-LTE approach in analogy to \citet{nie12}. This is
based on atmospheric structures calculated with 
the {\sc Atlas9}~code \citep{kur93b}. Line-formation calculations in
non-LTE were performed with recent versions of {\sc Detail} and {\sc Surface}
\citep[both updated by K.\,Butler;][]{gid81,but_gid85}.
The validity of the approach was
verified by direct comparison with hydrostatic and hydrodynamic line-blanketed
non-LTE model atmospheres \citep{nie07,ns11,prz11}.
Non-LTE level populations and synthetic spectra of all elements
were computed using updated model atoms as summarized in Table~2 of
\citet{prz16}.

An additional challenge for the analysis of mid-/late B-type stars as
discussed here are the overall reduced number of metal lines, which
are also typically much weaker than in the earlier B-type stars. 
Moreover, many of the mid-/late B-type stars show intermediate to large \vsini\ values,
such that often only a small number of metal lines remains for
the quantitative analysis besides the \ion{H}{i} and \ion{He}{i} lines. 
The only metal ionization equilibria available for the
atmospheric parameter determination (showing sensitivity to both
$T_\mathrm{eff}$ and $\log g$ variations) are those of \ion{O}{i/ii} and
\ion{Si}{ii/iii}. This renders the atmospheric parameter determination
less accurate than that of sharp-lined and hotter stars 
\citep[like those analyzed by][]{nie12}, for which the redundancy 
from additional ionization equilibria provides tighter constraints.
The \vsini\ values were derived from fits of the model spectra to
observed metal lines.

\begin{figure*}[htb]
\centering
\includegraphics[width=0.99\textwidth,clip]{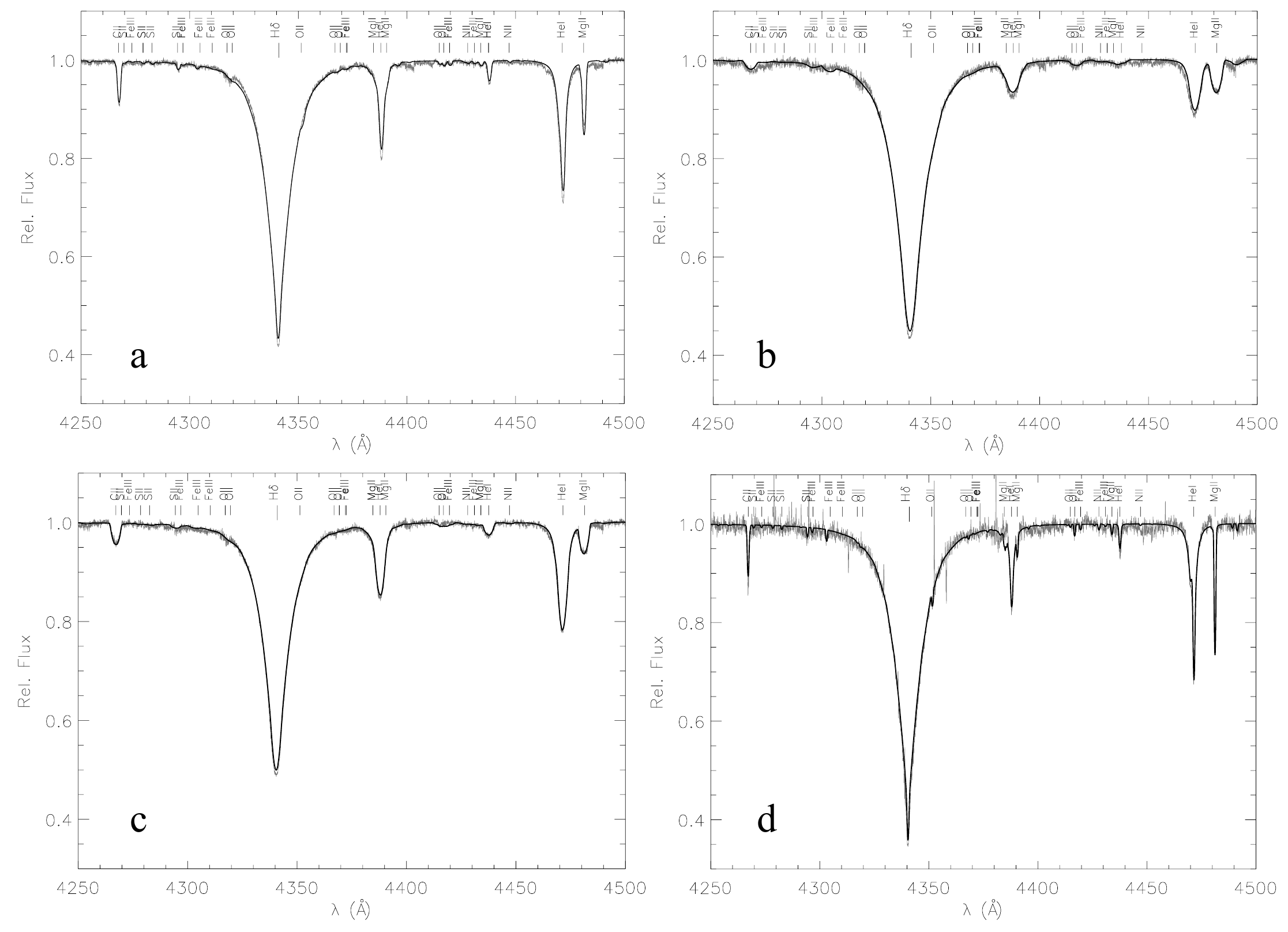}
\caption{Regions around the H$\gamma$ line for HD 47469 (panel a), HD 48012 (panel b), HD 261810 (panel c) and HD 261878 (panel d) showing the observations in dark grey and the fit with the corresponding synthetic spectra as black lines.}
\label{Hgamma}
\end{figure*}

The resulting atmospheric parameters of the four SPB type pulsators of
our star sample are summarized in Table~\ref{fundpars}.  
Examples for the quality of the fits to the observed spectra of the
four stars in the region of H$\gamma$ are shown in Figure~\ref{Hgamma}.
In the course of our analysis, we identified HD 47469 to be a binary system. We therefore list the atmospheric parameters of both components in Table \ref{fundpars}.
From the present data it is impossible to identify if one of the two components is pulsating or if both are SPB stars. 
The detailed analysis of the non-LTE quantitative spectral analysis
will be discussed in Paper~II.

\begin{table}[htb]
\caption{Fundamental parameters derived from spectroscopy and used for the asteroseismic analysis of the four SPB stars.} 
\label{fundpars}
\begin{center}
\begin{tabular}{lrlrr}
\hline
\hline
\multicolumn{1}{l}{\#} & \multicolumn{1}{c}{star} & \multicolumn{1}{c}{\Teff} 
& \multicolumn{1}{c}{log\,$g$} & \multicolumn{1}{c}{\vsini}    \\
\multicolumn{1}{l}{ } & \multicolumn{1}{c}{ }  & \multicolumn{1}{c}{[K]}  & \multicolumn{1}{c}{ } & \multicolumn{1}{c}{[kms$^{\rm -1}$]}     \\
\hline
1	&	HD 47469 A		&	17000 $\pm$ 500		&	4.25 $\pm$ 0.15		&	75 $\pm$ 15	\\
2	&	HD 47469 B		&	15000 $\pm$ 500		&	4.25 $\pm$ 0.15		&	125 $\pm$ 20	\\
3	&	HD 48012		&	14000 $\pm$ 500		&	4.30 $\pm$ 0.15		&	225 $\pm$ 20 \\
4	&	HD 261810		&	17200 $\pm$ 400		&	4.25 $\pm$ 0.10		&	180 $\pm$ 15 \\
5	& 	HD 261878		&	15200 $\pm$ 400		&	4.30 $\pm$ 0.10		&	40 $\pm$ 5 \\
\hline
\end{tabular}
\end{center}
\end{table}


\subsection{MESA Evolutionary Tracks}\label{ss-mesa}
We employed the state-of-the-art one-dimensional stellar structure and evolution code
MESA \citep[][version 7678]{pax11, pax13, pax15} with the
choice of physical input based on thorough asteroseismic inferences from the
modeling of two {\it Kepler} B8-type SPB stars KIC\,10526294 \citep{mor15} and 
KIC\,7760680 \citep{mor16}. 
We computed a coarse grid of non-rotating models by varying the initial mass M$_{\rm ini}$
from 4.0 to 8.0\,M$_\odot$, in steps of 0.5\,M$_\odot$.
MESA incorporates the effect of rotation on the equilibrium structure of stars by solving the modified structure equations presented by \cite{endal-1976-01,endal-1978-01}, including the effects of rotational mixing by several instabilities proposed by \cite{heger-2000-01}. The latter formalism contains a handful of unconstrained scaling coefficients which are non-trivial to constrain \citep[see e.g.][]{brott-2011-02}.
Considering that NGC\,2264 is 3 to 8 Myr old, the effect of rotation on redistributing the internal angular momentum in the target stars is negligible. Given the limited data at our disposal for each star, and the fact that the pulsation modes of these stars are unidentified, it is inconclusive to incorporate the effect of rotation on the structure; instead, we use non-rotating models, and assume the stars are uniformly rotating. The effect of rotation is only incorporated on the pulsations. For KIC\,7760680, \cite{mor16-03} demonstrated that differentially rotating models fail to reproduce the observed period spacing.

\begin{figure}
\includegraphics[width=\columnwidth]{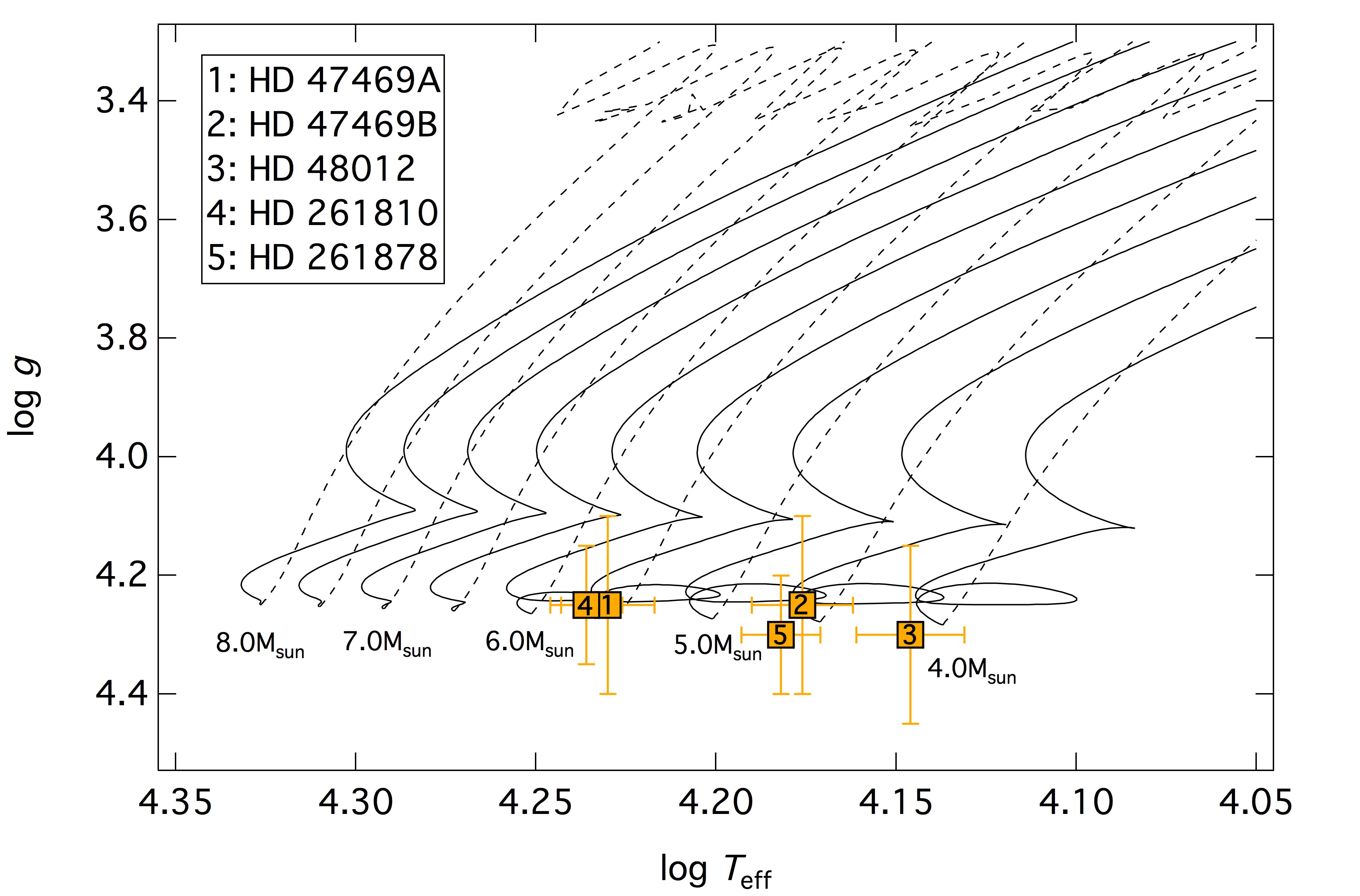}
\caption{Location of the four SPB type cluster members in the Kiel diagram.
Each star is given the number according to Table \ref{fundpars}. 
The pre-MS contraction phase of each track is shown with solid lines, and the main
sequence is marked by dotted lines for the range between 4 and 8\Msun\ in steps of 0.5\Msun.
The positions of the four SPB stars are so close to the ZAMS that even our asteroseismic modeling cannot 
discriminate whether or not they are pre-MS stars.
\label{f-kiel}} 
\end{figure}

The evolution of each model began from the pre-main sequence (pre-MS) phase, and was terminated as soon
as it reached the terminal age main sequence (TAMS, $X_{\rm core}=10^{-6}$).
We adopted the \citet{asp09} initial mixture with the initial composition 
$(X_{\rm ini}, Y_{\rm ini}, Z_{\rm ini})=(0.710, \,0.276, \,0.014)$, consistent with the Cosmic Galactic
Standard composition of \citet{nie12} for early-type stars in the solar neighbourhood.
We included convective core overshooting during the main sequence (MS) phase with the exponential overshoot
prescription of \citet{fre96} and \citet{her00}, which outperforms the classical 
step function prescription in reproducing the high precision g-mode frequencies of the above two \textit{Kepler} 
stars \citep[see e.g.,][for detailed comparison]{mor15,mor16} .
The free overshooting parameter was kept fixed at $f_{\rm ov}=0.025$.
For the same reason, we applied a minimum extra diffusive mixing in the radiative interior (beyond the 
overshoot region), with the value $D_{\rm ext}=100$ (cm$^2$\,s$^{-1}$), which is close to the two 
values found for KIC\,10526294 and KIC\,7760680. 
Along each evolutionary track, we stored an equilibrium model every $\sim100$\, K drop in effective
temperature $T_{\rm eff}$, provided $T_{\rm eff}$ of each model was in the range 11\,000 to 18\,000\,K.
The Kiel diagram (i.e., T$_{\rm eff}$ versus $\log g$; see Fig. \ref{f-kiel}) shows the pre-MS 
(solid) and main sequence  (MS, dotted) evolutionary tracks where the positions 
of the four NGC\,2264 SPB candidates derived from our non-LTE analysis are marked with the numbers given in Table \ref{fundpars}.

Clearly, all candidate stars lie close to the ZAMS.
The pre-MS evolution occurs on the Kelvin-Helmholtz contraction timescale, and the
MS evolution occurs on the nuclear timescale.
Given the large difference between these two timescales, the likelihood for these stars being on the pre-MS phase is slim.
Consequently, in the following seismic modeling, we adopt MS models to represent HD\,261810 and HD 47469.

using the spectroscopically determined fundamental parameters (see Table \ref{fundpars}; for more details see Paper II), with

Based on this grid, we found the closest model to the HR-diagram position of each star, using the MS part of evolutionary tracks to assess the model ages, accordingly.
Table\,\ref{t-ages} gives a summary of the inferred initial masses, ages and core hydrogen mass fraction X$_{\rm c}$ depending on the star being in the pre-MS or MS phase.
The inferred age spread is narrow, even with the ambiguity in the stars' burning status. 
The two components of the HD\,47469 system are expected to be coeval. From the 1$\sigma$ uncertainty of $T_{\rm eff}$ and $\log g$ of both components, the maximum possible age of the primary and secondary are 94 and 71 Myr, respectively. Therefore, the reported ages in Table\,\ref{t-ages} are agreeably close to one another.

The basic analysis presented above confirms previous assumptions that SPB pulsators that are clearly in their pre-MS phase are statistically hard to find as they evolve quite rapidly within a few million years through their pre-MS evolutionary stage. Even in a cluster as young as NGC 2264 the late B type objects that would be candidates for pre-MS SPB pulsation seem to already have started hydrogen core burning and have to be considered early ZAMS objects.

The available data for the young SPBs in NGC\,2264 suffer from short duration (of $<40$ days), and high noise level. Such data sets only allow a search for {\it indications} of $g$-mode period spacings. However, these time bases are insufficient for a more thorough asteroseismic analysis that would allow a study of possible structural differences (e.g., chemical gradients, mixing, convective core overshooting etc.) between young ZAMS and more evolved MS SPB stars. For the latter purpose, we would need period spacing series of at least a factor 2 longer that might be provided by future space missions such as TESS \citep{ric15} or PLATO \citep{rau14}.

\begin{table}
\caption{Inferred model masses and ages for candidate SPB targets in NGC\,2264, assuming they are in pre-MS or MS phase of their evolution.
The grid stepsize in mass is 0.5\,M$_\odot$.}
\label{t-ages}
\centering                          
\begin{tabular}{l l c c c c c c}        
\hline\hline                 
   &  &  & \multicolumn{2}{c}{PreMS} && \multicolumn{2}{c}{MS} \\ 
\cline{4-5}
\cline{7-8}      
\# & Star & M$_{\rm ini}$ & Age   & X$_{\rm c}$ && Age  & X$_{\rm c}$ \\
   & Name & [M$_\odot$]   & [Myr] &             && [Myr]&             \\
\hline  
1  & HD\,47469A & 5.5 & 1.10 & 0.71 && 5.68 & 0.69 \\ 
2  & HD\,47469B & 4.5 & 1.99 & 0.71 && 2.04 & 0.71 \\ 
3  & HD\,48012   & 4.0 & 2.71 & 0.71 && 2.78 & 0.71 \\ 
4 & HD\,261810  & 5.5 & 1.20 & 0.71 && 1.23 & 0.71 \\ 
5  & HD\,261878  & 4.5 & 1.99 & 0.71 && 2.04 & 0.71 \\ 
\hline  
\end{tabular}
\end{table}

\subsection{Period spacings in the SPB stars}
In rotating stars, the pulsation eigenfrequeincies in the co-rotating frame are sensitive to rotation \citep{unno-1989-book}. The period spacing of zonal and prograde modes decreases monotonically with increasing rotation velocity \citep{tow03,bouabid2013}.
Using the results of our frequency analysis for the four SPB stars HD 47469, HD 48012, HD 261810 and HD 261878, we investigated the presence of regular gravity (g) mode period spacings and the deviations from their averages. We inspected the period content for each of the stars separately and computed the corresponding period spacings. Because the time bases of the MOST observations are rather short ($\sim$ 24 and 39 days), our search for nearly equidistant period spacings is limited, and does not allow us to investigate rotational splittings of g-modes.
From the four SPB stars in NGC 2264, we find indications for period spacing patterns in two stars: HD 261810 and HD 48012.

\begin{figure}[htb]
\centering
\includegraphics[width=0.95\columnwidth,clip]{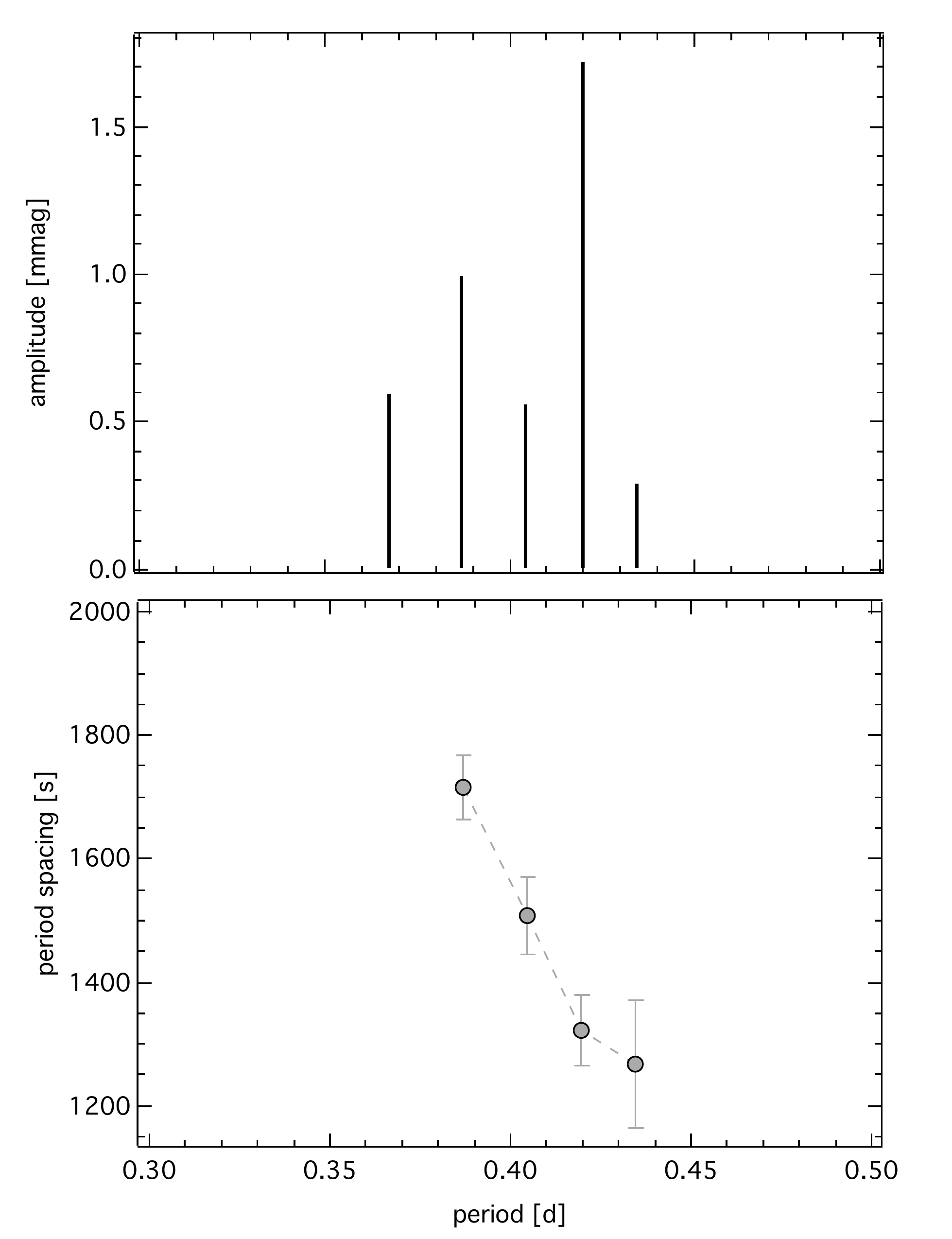}
\caption{The observed period spacings in HD 261810.}
\label{HD261810_Pdiff}
\end{figure}

\begin{figure}[htb]
\centering
\includegraphics[width=0.95\columnwidth,clip]{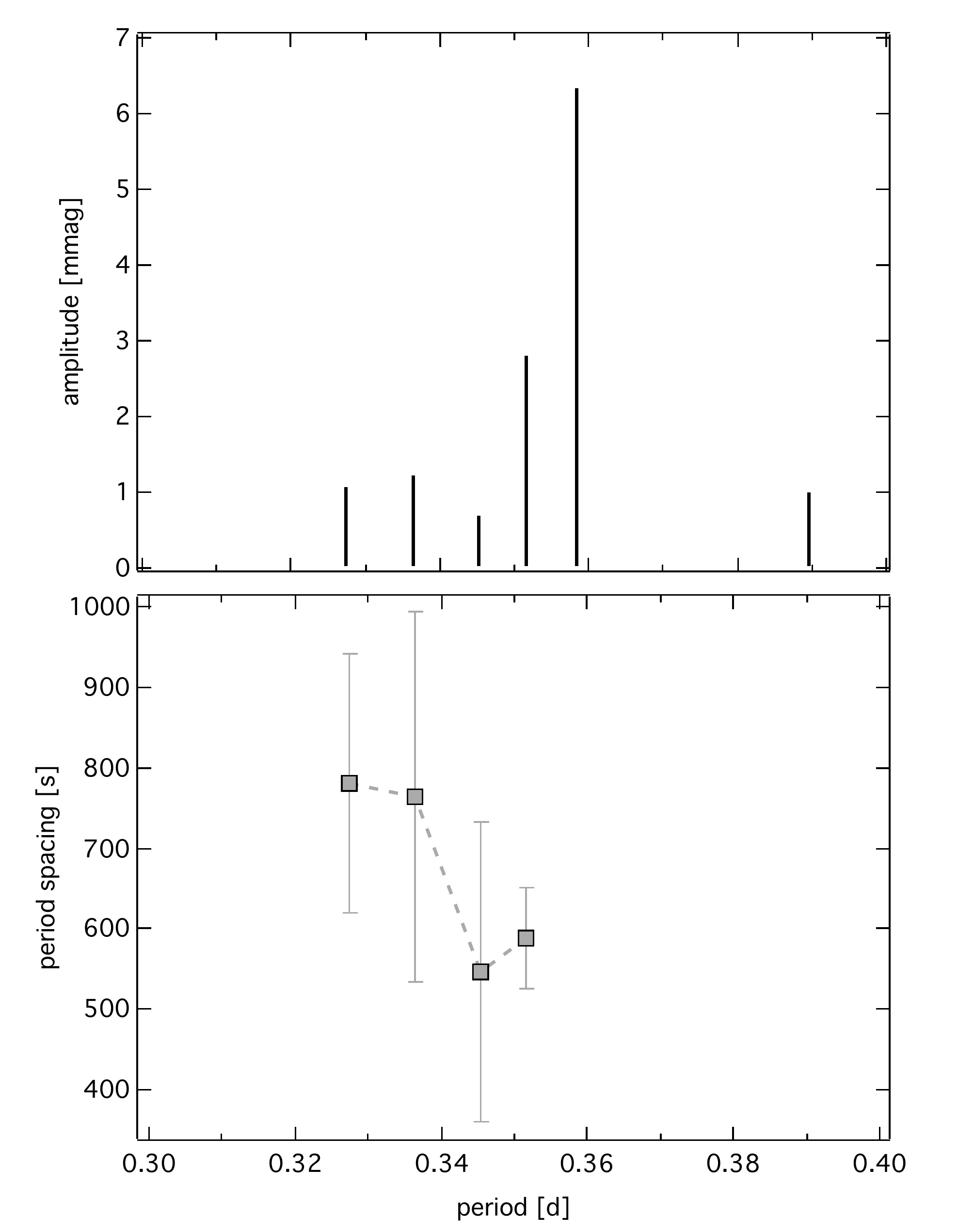}
\caption{The observed period spacings in HD 48012.}
\label{HD48012_Pdiff}
\end{figure}

The star HD 261810 exhibits a regular period spacing structure (see Fig. \ref{HD261810_Pdiff}), assuming that all modes belong to a series with identical spherical degree $\ell$. The longest series we find includes five consecutive peaks, but the similar spacing can also be found between two smaller periods (around 0.2 days).
The average spacing is 1495 seconds (see Fig. \ref{HD261810_Pdiff}) with a maximum absolute deviation of 230 seconds.

The pulsational periods found in HD 48012 also show an average spacing of 669 seconds with a maximum absolute deviation of 120 seconds which is illustrated in Fig. \ref{HD48012_Pdiff}.

\begin{figure*}[htb]
\centering
\includegraphics[width=0.9\textwidth,clip]{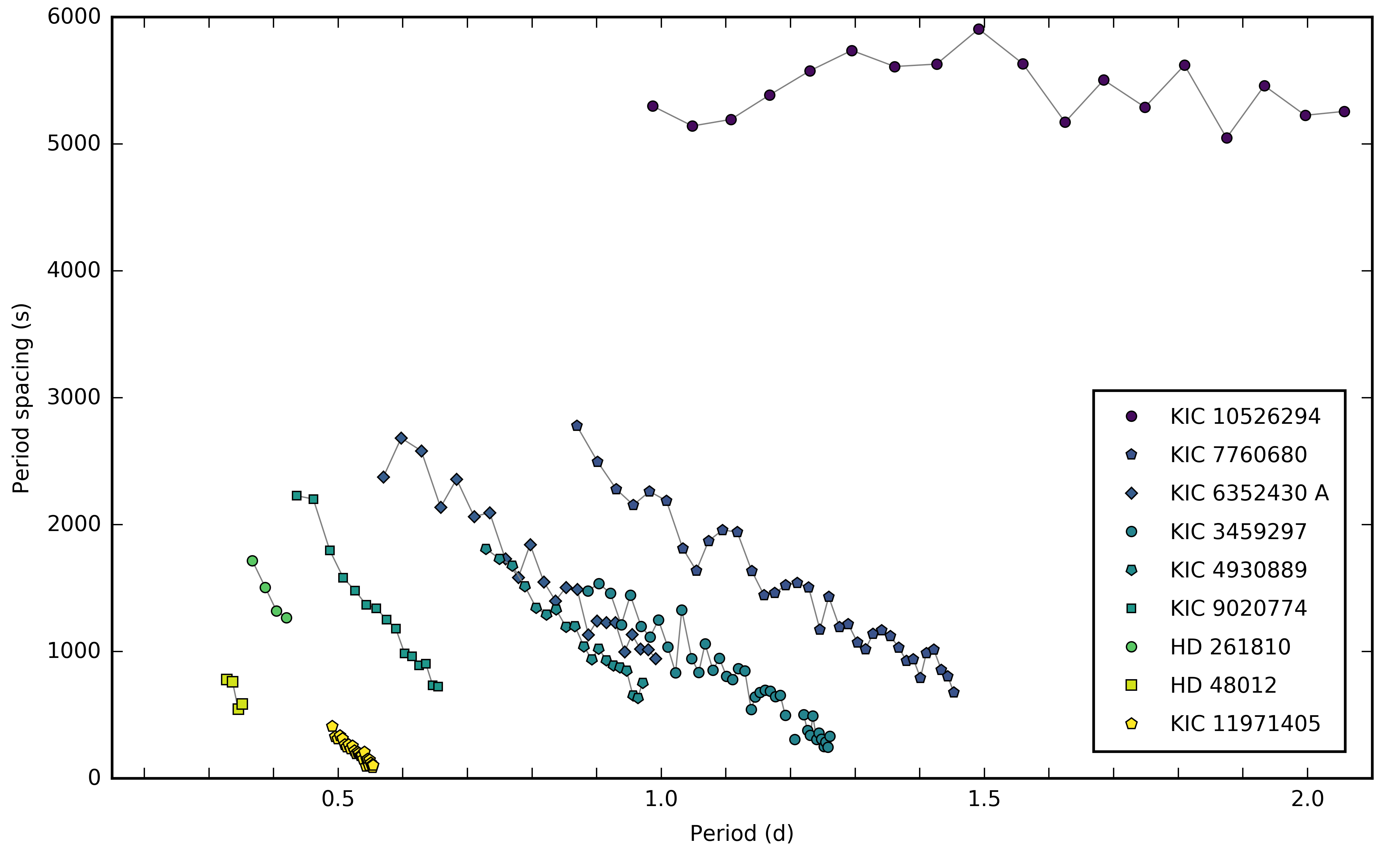}
\caption{Comparison of the period spacing patterns of HD 48012 (yellow squares) and HD 261810 (green circles) to the seven SPB stars observed by the \textit{Kepler} space mission and discussed in \citet{pap16}.}
\label{Keplercomparison}
\end{figure*}

Figure \ref{Keplercomparison} shows a comparison between the short period spacing patterns identified from the MOST data of HD 48012 (yellow squares) and HD 261810 (yellow circles) and seven stars observed by the Kepler mission \citep[grey squares,][]{pap16}. The authors discuss that stars with higher rotational velocity show their strongest excited modes and their main period series towards shorter periods, and their average spacing values are smaller. Our two young SPB stars, HD 48012 and HD 261810, with \vsini\, values of 225$\pm$20 \kms\ and 180$\pm$15 \kms\, follow that relation nicely as they belong to the fastest rotators investigated, they show the shortest periods and smallest spacing values.

No clear spacing pattern can be identified from the 12 pulsation periods found in the HD 47469 data. The reasons might also be connected to the fact that HD 47469 was discovered to be a spectroscopic binary (see Paper II) that might complicate the pulsational analysis to the extent that both components might be pulsating.  

The present data for HD 261878 are too sparse to calculate an average observed spacing and its deviations. It seems that periods at lower amplitudes might be buried in the noise and could fill the gaps between the six periods identified by us. It is therefore obvious that longer photometric time series with lower noise levels are necessary to recover the possibly missing periods and reveal the complete period spacing pattern.

\subsection{Asteroseismic Inference for HD\,261810}

As explained in Sect.\,\ref{star4}, HD\,261810 exhibits five g-modes with nearly equidistant period spacing (F1, F4, F5, F6, and F10 in Table\,\ref{hd261810freqs}). 
From Fig.\,\ref{f-kiel}, it can be seen that HD\,261810 is very close to the zero-age main sequence (ZAMS), with \vsini\ = 180$\pm15$ km\,s$^{-1}$.From the light curve analysis, it is impossible to identify the angular quantum numbers $(\ell,m)$, since there are no reconcilable frequency splittings around each peak (see Fig.~\ref{hd261810ampspec}). For high rotation velocities, rotational splittings will be large and asymmetric.

To remedy this, we benefit from the available seismic information, i.e., the periods of the observed five g-modes explained above, and examine if a seismic analysis can help identifying the observed modes. We computed mode stability analysis with the state-of-the-art linear nonradial nonadiabatic one-dimensional oscillation code GYRE \citep[v.\,4.4,][]{tow13}. GYRE accounts for the effect of rotation on pulsation eigensolutions by employing the traditional approximation of rotation \citep{eck60}, which is accomplished by ignoring the  $\theta$-component of angular velocity in the governing oscillation equations \citep{lee86,lee97,tow03}. This is a valid approach, because \citet{tow05a,tow05b} has already shown that the coupling of the $\kappa$-mechanism and the Coriolis force provides a sufficient restoring force for the excitation of gravito-inertial waves in rotating SPB stars. Two confirmed examples of gravito-inertial waves in B stars are HD\,43317 \citep{2012A&A...542A..55P} and KIC\,7760680 \citep{mor16}. 

As an input to GYRE, we use a MESA model that is closest to the position of HD\,261810 on the Kiel diagram (Fig.\,\ref{f-kiel}), and has just started burning hydrogen in the core.
Given the projected rotational velocity \vsini\ = 180$\pm15$ km s$^{-1}$ , and the five regularly spaced periods, we attempt to reproduce the observed modes.
To that end, we vary the mode spherical degrees $(\ell,\,m)$ and the uniform rotation
$\Omega_{\rm rot}$. At the same time, we try to satisfy the following five constraints regarding the observed frequencies: 
(a) they must remain positive-valued in the co-rotating frame,
(b) they must be predominantly excited,
(c) they must be consecutive in radial order, as inferred from observations,
(d) the equatorial rotation velocity must exceed the projected rotation velocity, i.e. $v_{\rm rot}\geq v\sin i$ to ensure the projection factor of the inclination angle stays between zero and unity $|\sin i|\leq 1$, and
(e) the observed g-modes are zonal or prograde $m\geq0$, given the negative slope of the period spacing (bottom right panel of Fig.\,\ref{Keplercomparison}).
The latter constraint stems from the theoretical predictions of \cite{tow03} and \cite{bouabid2013}, and is confirmed by the observational and theoretical modeling of
KIC\,7760680 \citep{pap15,mor16} and intermediate mass g mode $\gamma$ Dor pulsators \citep{van16}.
From flux cancellation of unresolved stellar disks, we know that the dipole ($\ell=1$) and quadrupole ($\ell=2$) modes have greater chance to be observed in white-light photometry than higher-degree modes do \citep{aer10}.
For these reasons, we restrict our further seismic analysis to dipole ($\ell=1$) and quadrupole ($\ell=2$) zonal and prograde g-modes.

\begin{figure*}
\begin{minipage}{0.48\textwidth}
\includegraphics[width=\columnwidth]{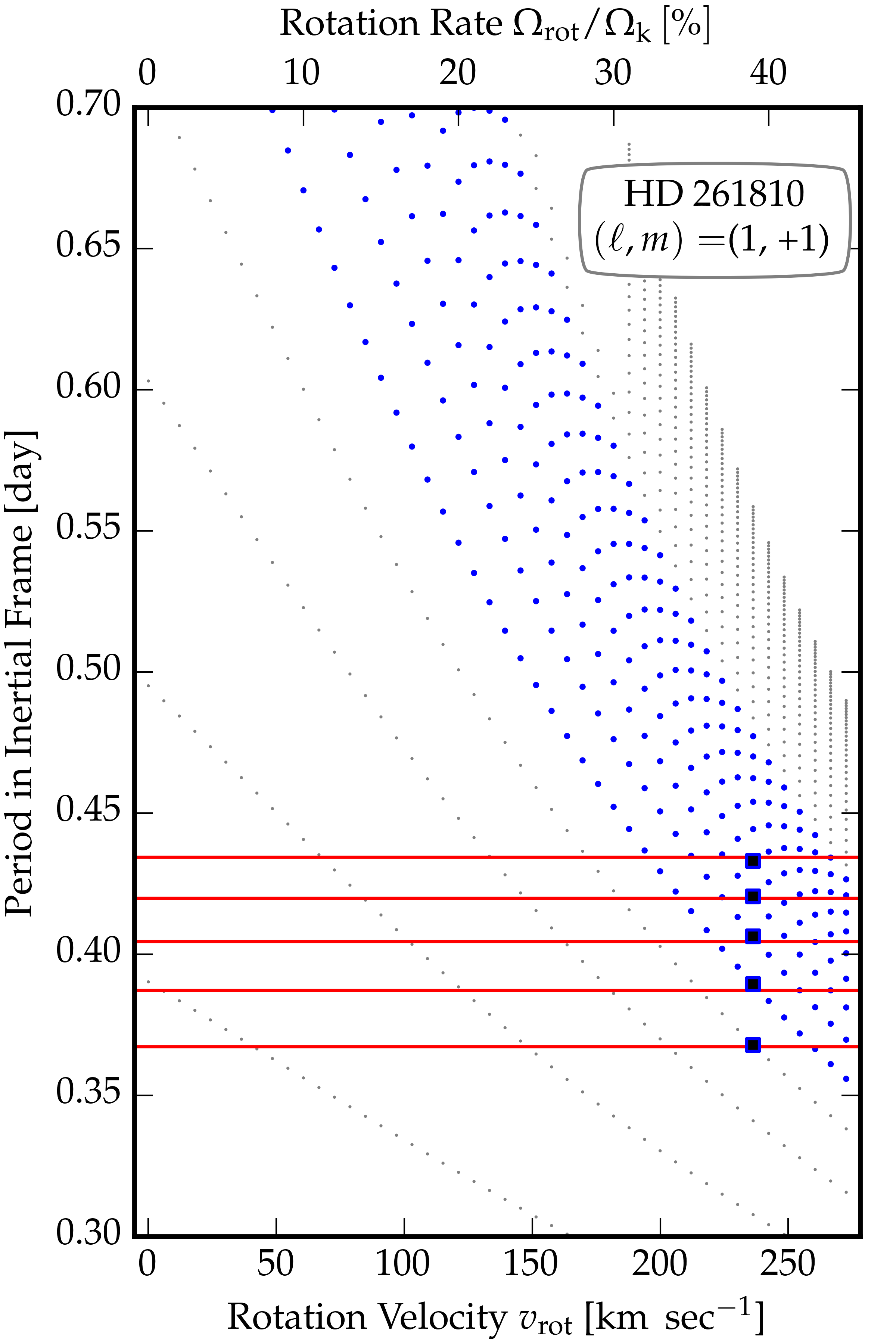}
\end{minipage}
\begin{minipage}{0.48\textwidth}
\includegraphics[width=\columnwidth]{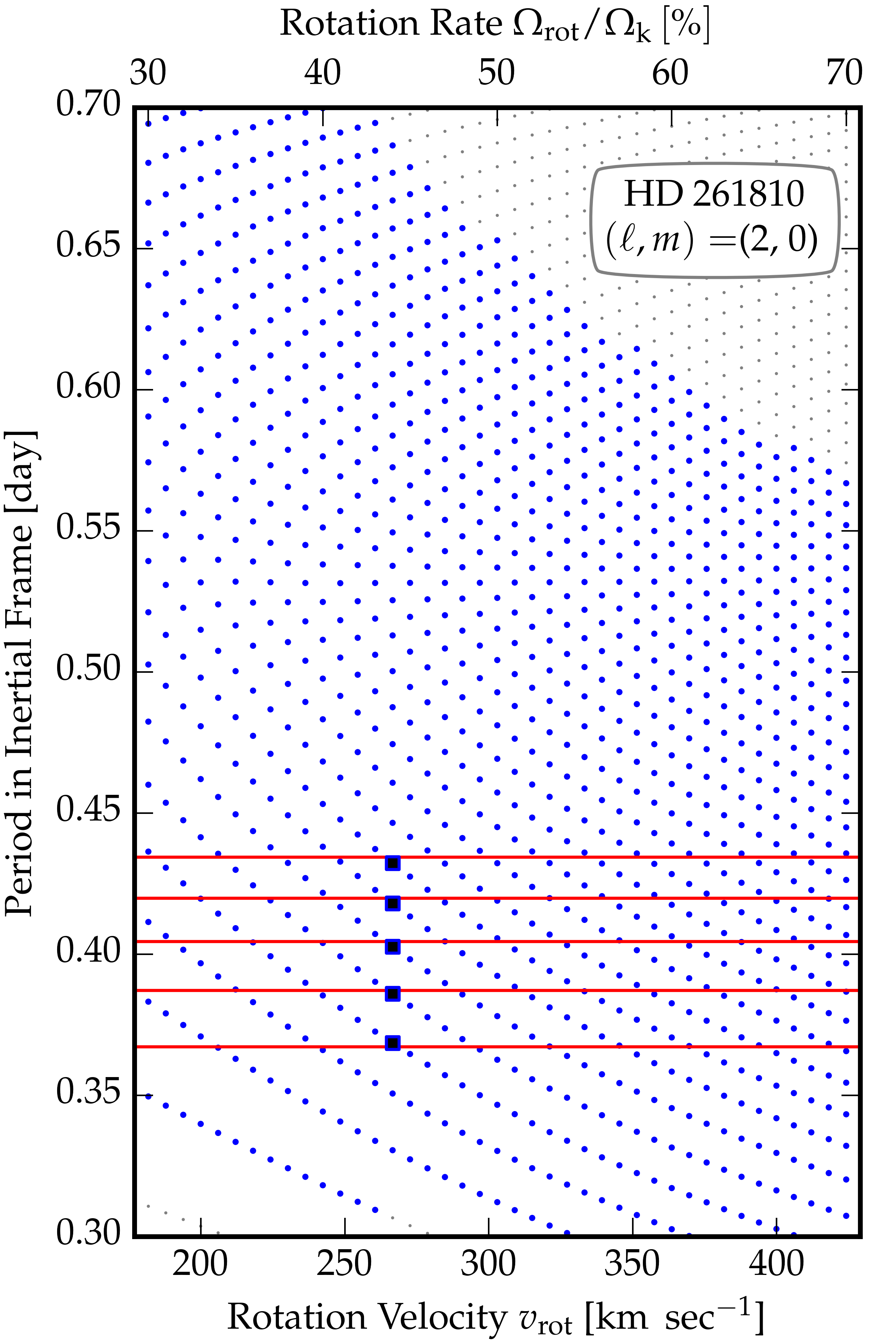}
\end{minipage}
\caption{Evolution of mode periods in the inertial frame versus equatorial rotation velocity. The observed periods of HD\,261810 are shown with horizontal red lines. The stables modes are shown with grey dots, while the excited modes are shown with blue dots. {\it Left:} The dipole prograde modes provide a possible period match for $v_{\rm rot}=235$ km\,s$^{-1}$. {\it Right:} The quadrupole zonal modes can equally provide a fit for $v_{\rm rot}=270$ km\,s$^{-1}$. Table\,\ref{t-modes} summarizes the results for different $\ell$ and $m$.}
\label{f-el-1-2}
\end{figure*}

\cite{tow03} and \cite{bouabid2013} have already demonstrated that for a fixed harmonic degree $\ell$ and uniform rotation velocity, the g-mode period spacing for tesseral prograde modes $m>0$ has a steeper slope than that of zonal $m=0$ modes. In other words, to reproduce the observed period spacing slope of HD\,261810, we can either adopt much higher rotational velocity for zonal modes, or adopt a lower rotational velocity for tesseral modes. In the following, we adopt $(\ell,\,m)=(1,\,0)$, (1,\,$+1$), (2,\,0), (2,\,$+1$) and (2,\,$+2$), and try to simultaneously fulfill the five constraints (a) to (e) given above.

\begin{table}
\centering
\caption{Different possibilities for the mode harmonic degrees ($\ell, \,m$)
for HD\,261810.
}\label{t-modes}
\begin{tabular}{lccccl}
\hline \hline
$(\ell, \,m)$ & $v_{\rm rot}$ & $\Omega_{\rm rot}/\Omega_{\rm K}$ & $v_{\rm rot}/v \sin i$ & $i$ & Accept? \\
& (km/s) & (\%) & & (deg) & (Yes/No) \\
\hline
(1, \,0) & --  & -- & -- & -- & No \\
(1, \,1) & 236 & 39 & 1.3 & 48 & Yes \\
(2, \,0) & 267 & 44 & 1.5 & 41 & Yes \\
(2, \,1) & 135 & 22 & 0.75 & -- & No \\
(2, \,2) & 82  & 14 & 0.45 & -- & No \\
\hline
\end{tabular}
\end{table}

For each choice of $\ell$ and $m$, we search for a rotation rate for which the five observed periods coincide with five consecutive g-modes. Our findings are summarized in Table\,\ref{t-modes}. The first column gives the choice of spherical harmonics, and the second column gives the optimal rotation velocity $v_{\rm rot}$ from a model that sufficiently matches the observed periods. The third column gives the corresponding rotation rate $\Omega_{\rm rot}/\Omega_{\rm K}$ in percent. The Keplerian break up velocity is $\Omega_{\rm K}=\left(GM_\star/R_\star^3\right)^{1/2}$. The fourth column gives the ratio of the rotation velocity to the projected rotation velocity $v_{\rm rot}/v \sin i$, where $i$ is the inclination angle in degrees. The last column concludes if the corresponding harmonic degrees can provide a possible explanation to the observed modes.

Figure\,\ref{f-el-1-2} shows the evolution of predicted excited (blue) versus damped (grey) mode 
periods in the inertial (observer's) frame versus the equatorial rotation velocity on the lower abscissa. The upper abscissa shows $\Omega_{\rm rot}/\Omega_{\rm K}$. The horizontal red lines mark the five observed periods of HD\,261810.
The dipole prograde $(\ell,\,m)=(1,\,1)$ and quadrupole zonal $(\ell,\,m)=(2,\,0)$ modes are shown on the left (and right) panels. The best matching modes are shown with larger blue squares.
Similar figures for other spherical degrees are not presented here for brevity.

The dipole zonal modes cannot explain the observed modes, unless the star rotates $\gtrsim$85\% its break up rotation frequency.
At such rotation rates, B stars are expected to form a decretion disk, turn into B[e]-type stars, and exhibit emission spectral lines \citep[e.g.,][]{riv13}. However, this contradicts the observed spectrum of this star (Paper II), which does not show emission lines. Thus, we interpret this as an unlikely circumstance, and exclude dipole zonal modes. The quadrupole tesseral modes provide a possible match to the observed periods, but only for rotational velocities significantly smaller than the projected rotational velocity; as the latter contradicts with the constraint (d), we exclude quadrupole tesseral modes, too.

The dipole prograde and quadrupole zonal modes provide a degenerate solution to the observations. They both equally fulfill all modeling constraints (a) to (e). The shortest-period dipole prograde mode that fits the observations is found to be stable. This is not a major concern since a slight increase in metallicity can increase the height of the iron opacity bump, and destabilize the additional mode. For young B-stars, a slight metallicity excess is justifiable. Despite the fact that the two above models are degenerate, the valid range of the rotational velocity and the inclination angle are narrow. The two plausible rotational velocities, 236 or 267 km\,s$^{-1}$, are only $\sim$12\% different. The two inclination angles of 41 or 48 degrees imply that HD\,261810 is observed at an intermediate inclination angle which keeps both sets of modes visible.

\subsection{Asteroseismic Inference for HD\,48012}

The star HD\,48012 has the highest \vsini\, of the sample with \vsini = 225$\pm$25\,km\,s$^{-1}$. It exhibits six significant pulsation frequencies, five of which form a period spacing series with a negative slope (Fig\,\ref{HD48012_Pdiff}). In this regard, the same analysis that was already presented in the previous section can be repeated for this star. 

\begin{figure*}
\begin{minipage}{0.48\textwidth}
\includegraphics[width=\columnwidth]{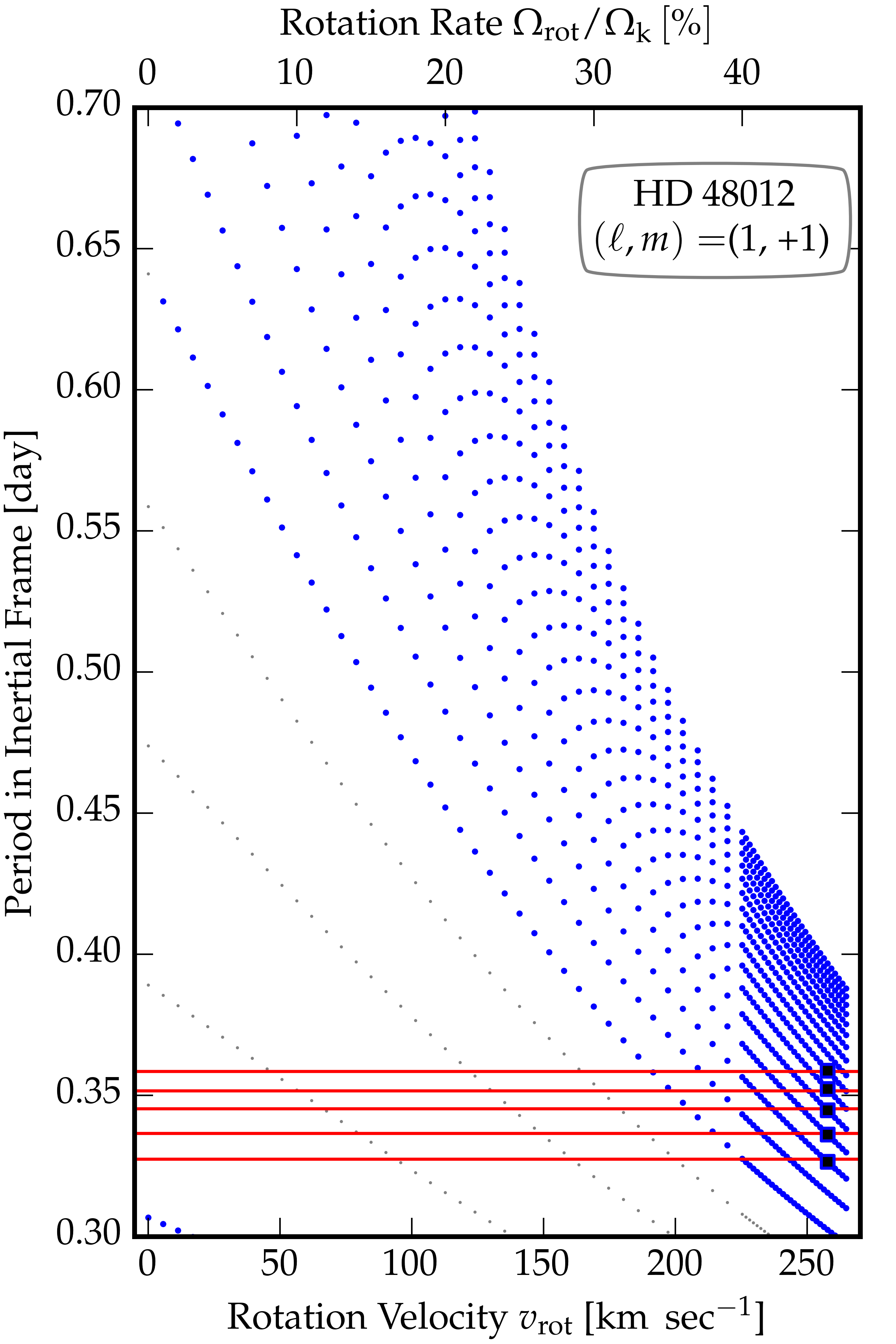}
\end{minipage}
\begin{minipage}{0.48\textwidth}
\includegraphics[width=\columnwidth]{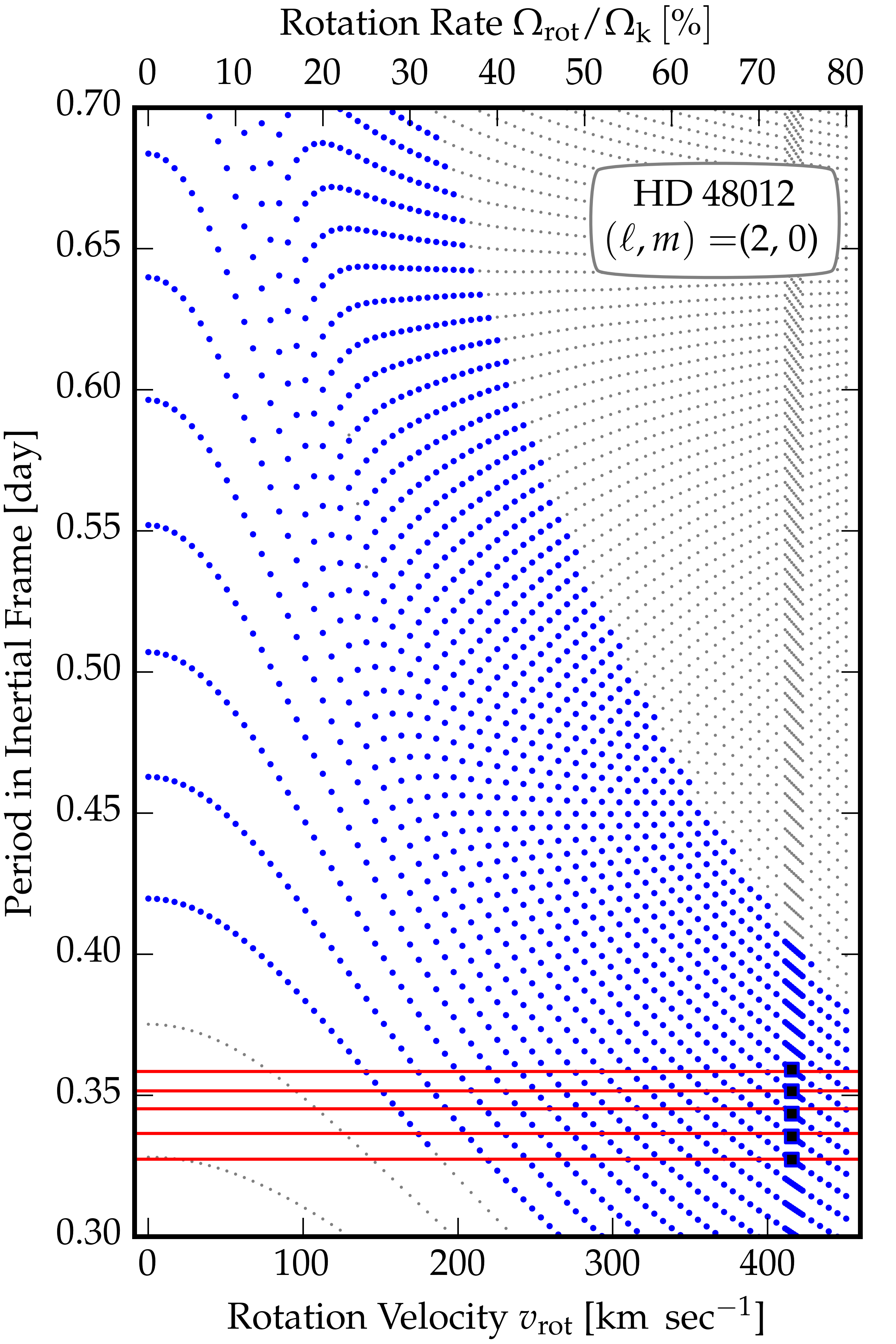}
\end{minipage}
\caption{Similar to Fig.\,\ref{f-el-1-2} for HD\,48012. Note that the projected rotation velocity is $v\sin i=220\pm20$ km s$^{-1}$.
The $(\ell,\,m)=(2,\,0)$ is rejected because it requires high rotation velocities, giving rise to B[e] phenomenon, which is not observed in spectroscopy.} 
\label{f-seism-hd48012}
\end{figure*}

Table\,\ref{t-modes-2} and Fig.\,\ref{f-seism-hd48012} present the modeling results. The dipole zonal modes cannot represent the observed periods because they exhibit no matching patterns even up to 80\% break up velocity. The same is true for the quadrupole tesseral modes. The dipole prograde modes can reproduce the observed periods (Fig.\,\ref{f-seism-hd48012} left panel), provided the star rotates uniformly at 45.75\% break up velocity, or with an equatorial velocity of 258 km s$^{-1}$. In this case, the inclination angle is 61 degrees. Alternatively, the quadrupole zonal modes can also provide a fit to the observed periods (Fig.\,\ref{f-seism-hd48012} right panel), provided the star rotates uniformly at 74\% break up velocity. This corresponds to an equatorial rotational velocity of 416 km\,s$^{-1}$ with an inclination angle of 33 degrees. Similar to the previous star, the fast rotation rate deduced from quadrupole zonal modes is unlikely, because there is no signature of B[e] phenomenon in the spectrum of this star. Consequently, we reject this identification for the modes.

The validity domain of the traditional approximation, compared to the results of two-dimensional frequency computations is limited to $\sim40\%$ \citep{ballot-2012-01,prat-2016-02,ouazzani-2016-01}. The proposed solution for HD\,48012 (Table \ref{t-modes-2}) exceeds this validity domain. To a much lesser extent, the same is true for HD\,261810. Consequently, fitting the observed periods of HD\,48012 with quadrupole zonal modes -- which requires $\Omega_{\rm rot}/\Omega_{\rm K}=74\%$ -- is unreliable. Based on this, the corresponding rotational velocity and inclination angles are subject to large model uncertainties.

\begin{table}
\centering
\caption{Different possibilities for the mode harmonic degrees ($\ell, \,m$)
for HD\,48012.
}\label{t-modes-2}
\begin{tabular}{lccccl}
\hline \hline
$(\ell, \,m)$ & $v_{\rm rot}$ & $\Omega_{\rm rot}/\Omega_{\rm K}$ & $v_{\rm rot}/v \sin i$ & $i$ & Accept? \\
& (km/s) & (\%) & & (deg) & (Yes/No) \\
\hline
(1, \,0) & --  & -- & -- & -- & No \\
(1, \,1) & 258 & 46 & 1.15 & 61 & Yes \\
(2, \,0) & 416 & 74 & 1.85 & 33 & No \\
(2, \,1) & -- & -- & -- & -- & No \\
(2, \,2) & --  & -- & -- & -- & No \\
\hline
\end{tabular}
\end{table}

\section{Summary and conclusions}

We have conducted a first comprehensive investigation of the variability of B stars in the young open cluster NGC 2264 using space photometry obtained by the MOST, CoRoT and Spitzer satellites. This will be complemented by a detailed study of the fundamental parameters, chemical abundances, presence of magnetic fields, spectral energy distributions and distances in the second part of this work (i.e., Paper II by Przybilla et al. 2017, in prep.).
Our sample comprises in total eleven stars in the field of the cluster where one object, HD 261054, was identified as non-member with Be type variability and the remaining ten stars are likely cluster members according to their proper motions, radial velocities, distances and positions in the cluster colour-magnitude diagram. These ten members of NGC 2264 show variability due to SPB type pulsations (HD 47469, HD 48012, HD 261810, HD 261878), due to rotational modulation caused by spots on their surfaces (HD 47777, HD 47887, HD 261903, NGC 2264 137, and possibly HD 261938) or are connected to binarity (HD 47961 and possibly HD 261938) which will be discussed in detail in Paper II.

The WEBDA database of open clusters\footnote{http://webda.physics.muni.cz} lists 32 B type stars in the field of NGC 2264; twenty-six of these are cluster members according to their proper motions \citep{vas65}, their radial velocities \citep{fos14} or identified to be likely members in our study. The ten variable stars analyzed here are included in this list, and therefore correspond to a percentage of 38\% variable and 15\% pulsating stars among the spectral type B objects in NGC 2264. 

Originally, we intended to search for pre-MS SPB type stars in NGC 2264. Using our theoretical models computed with the MESA stellar structure and evolution code \citep{pax11,pax13,pax15} in combination with the stars' fundamental parameters derived from high-resolution spectroscopy (Paper II), we find them to be located near the ZAMS in the HR-diagram (Fig. \ref{f-kiel}). Although their ages computed from these models lie between only $\sim$ 1 and 6 million years, all stars seem to be early ZAMS objects that are not in their pre-MS phase any more. This illustrates how fast early stellar evolution progresses for (late) B type stars and that bona-fide pre-MS SPB stars are statistically hard to discover.

Two of the four SPB type pulsators show regularly spaced g-mode period series which we used for a first asteroseismic analysis. For HD 261810 two possible mode identifications are either ($\ell$,$m$) = (1,+1) or ($\ell$,$m$) = (2,0). For HD 48012, the only plausible solution is ($\ell$,$m$) = (1,+1). We also compare the period spacings of our two stars (HD 48012 and HD 261810) to those of seven SPB stars observed by the Kepler main mission \citep{bor10} and analyzed by \cite{pap16}. With projected rotational velocities, \vsini, of 225$\pm$25 and 180$\pm$15\kms\ for HD 48012 and HD 261810, respectively, they are among the fastest rotators in that sequence, and, hence, show the shortest periods and period spacings. 

Both HD\,261810 and HD\,48012 are fast rotators, with their $\Omega_{\rm rot}$ exceeding $\sim40\%$ of their breakup rotation frequency (see Tables \ref{t-modes} and \ref{t-modes-2}). For such high rotation rates, the validity of traditional approximation of rotation is questionable \citep{mirouh-2016-01,prat-2016-02}. However, those formalisms were developed based on polytropic models, and the effect of chemical stratification on the pulsational behavior was ignored. As a result, our findings regarding the rotational velocity and inferred inclination angles might change if one goes beyond the traditional approximation of rotation, or uses 2D pulsation codes \citep{ouazzani-2016-01}. Consequently, these two stars are ideal test cases for more sophisticated models of rotating pulsating B stars, if higher-precision data with longer time baseline is collected in the future with e.g. the PLATO 2.0 mission.

The currently available photometric time series for the young B stars have time bases of less than 40 days. Hence, the noise level is too high to detect longer period spacing series for HD 48012 and HD 261810 and to discover possible period spacing patterns for HD 47469 and HD 261878. With longer photometric time series, e.g., from future space missions like TESS \citep{ric15} and PLATO \citep{rau14}, we will also be able to conduct a more in-depth asteroseismic analysis of the internal structures of young ZAMS B type pulsators and compare them to their more evolved counterparts. With such data we will also be able to continue our search for bona fide pre-MS SPB stars.

\begin{acknowledgements}

This work includes observations made with the {\it Spitzer} Space Telescope. {\it Spitzer} is operated by the Jet Propulsion Laboratory, California Institute of Technology under a contract with NASA. Support for this work was provided by NASA through an award issued by JPL/Caltech.
This research has made use of the NASA/ IPAC Infrared Science Archive, which is operated by the Jet Propulsion Laboratory, California Institute of Technology, under contract with the National Aeronautics and Space Administration, and of the WEBDA database of open clusters originally developed by J.-C. Mermilliod (Laboratory of Astrophysics of the EPFL, Switzerland), operated and maintained by E. Paunzen, C. St\"utz and J. Janik at the Department of Theoretical Physics and Astrophysics of the Masaryk University, Brno (Czech republic).
This work was also conducted using data from the European Space Agency (ESA)
mission {\it Gaia} (\url{http://www.cosmos.esa.int/gaia}), processed by
the {\it Gaia} Data Processing and Analysis Consortium (DPAC,
\url{http://www.cosmos.esa.int/web/gaia/dpac/consortium}). Funding
for the DPAC has been provided by national institutions, in particular
the institutions participating in the {\it Gaia} Multilateral Agreement.
\\
KZ acknowledges support by the Austrian Fonds zur F\"orderung der wissenschaftlichen Forschung (FWF, project V431-NBL).
PIP has received funding from The Research Foundation -- Flanders (FWO), Belgium. EM has received funding from the People Programme (Marie Curie Actions) of the European Union's Seventh Framework Programme FP7/2007-2013 under REA grant agreement No. 623303 (ASAMBA). The research leading to these results has (partially) received funding from the European Research Council (ERC) under the European Union's Horizon 2020 research and innovation programme (grant agreement N$^{\circ}$670519: MAMSIE).
MFN acknowledges support by the Austrian Austrian Fonds zur F\"orderung der wissenschaftlichen Forschung (FWF) in the form of a Meitner Fellowship (project N-1868-NBL).
RK is supported by the Austrian Research Promotion Agency - ALR. 
Funding for the Stellar Astrophysics Centre is provided by The Danish National Research Foundation (Grant DNRF106). The research was supported by the ASTERISK project (ASTERoseismic Investigations with SONG and Kepler) funded by the European Research Council (Grant agreement no.: 267864).
NT has received funding from the European Research Council under the European Community's Seventh Framework Programme (FP7/2007-2013) / ERC grant agreement no 338251 (StellarAges).
\\
We would like to thank T. Ryabchikova and Y. Pakhomov for their valuable input and fruitful discussions about this work.
\end{acknowledgements}

\bibliographystyle{aa} 
\bibliography{NGC2264Bstars_phot.bib} 

\end{document}